\documentclass[journal,twocolumn]{IEEEtran}
\usepackage{amsfonts}
\usepackage{times}
\usepackage{graphicx}
\usepackage{latexsym}
\usepackage{dsfont}
\usepackage{amssymb}
\usepackage{amsmath}
\usepackage{cite}
\usepackage{verbatim}

\newcommand{\figref}[1]{{Fig.}~\ref{#1}}

\def\bb0{{\mathbb{0}}}

\def\ba{{\mathbf{a}}}
\def\bb{{\mathbf{b}}}

\def\b0{{\mathbf{0}}}

\def\sf0{{\mathsf{0}}}

\usepackage{epstopdf}
\usepackage{enumerate}
\usepackage{algorithmicx}
\usepackage{algorithm}
\usepackage{amsmath}
\usepackage{mathtools}
\usepackage[noend]{algpseudocode}
\usepackage{float}
\usepackage{hyperref}
\usepackage{color}
\usepackage{makeidx}
\usepackage{bbm}
\usepackage{graphicx}
\usepackage{lipsum}
\usepackage{subcaption}
\usepackage{tablefootnote}
\usepackage{multirow}
\usepackage{multicol}
\usepackage{balance}
\usepackage{booktabs}
\usepackage{soul}
\usepackage{xcolor}
\usepackage{autobreak}
\usepackage[utf8]{inputenc}

\usepackage[font=scriptsize]{caption}

\def\j{\mathrm{j}}

\newcommand{\comm}[1]{}

\begin{document}

\title{Camera Based mmWave Beam Prediction:  \\ Towards Multi-Candidate  Real-World Scenarios}
\author{Gouranga Charan, Muhammad Alrabeiah, Tawfik Osman, and Ahmed Alkhateeb \thanks{The authors are with the School of Electrical, Computer, and Energy Engineering, Arizona State University. Emails: \{gcharan, malrabei, tosman, alkhateeb\}@asu.edu. Muhammad Alrabeiah is currently with the Electrical Engineering Department, King Saud University, Riyadh 11421, Saudi Arabia. This work was conducted while he was with the Wireless Intelligence Lab at the School of ECEE, ASU, Tempe, AZ, USA.  This work was supported by the National Science Foundation (NSF) under Grant No. 2048021.}}

\maketitle

\begin{abstract}
Leveraging sensory information to aid the millimeter-wave (mmWave) and sub-terahertz (sub-THz) beam selection process is attracting increasing interest. This sensory data, captured for example by cameras at the basestations, has the potential of significantly reducing the beam sweeping overhead and enabling highly-mobile applications. The solutions developed so far, however, have mainly considered single-candidate scenarios, i.e., scenarios with a single candidate user in the visual scene, and were evaluated using synthetic datasets. To address these limitations,  this paper extensively  investigates the sensing-aided beam prediction problem in a real-world multi-object vehicle-to-infrastructure (V2I) scenario and presents a comprehensive machine learning based framework. In particular, this paper proposes to utilize visual and positional data to predict the optimal beam indices as an alternative to the conventional beam sweeping approaches. For this, a novel user (transmitter) identification solution has been developed, a key step in realizing sensing-aided multi-candidate and multi-user beam prediction solutions. The proposed solutions are evaluated on the large-scale real-world DeepSense $6$G dataset. Experimental results in realistic V2I communication scenarios indicate that the proposed solutions achieve close to $100\%$ top-5 beam prediction accuracy for the scenarios with single-user and close to $95\%$ top-5 beam prediction accuracy for multi-candidate scenarios. Furthermore, the proposed approach can identify the probable transmitting candidate with more than $93\%$ accuracy across the different scenarios. This highlights a promising approach for nearly eliminating the beam training overhead in mmWave/THz communication systems.  

\end{abstract}


\section{Introduction} \label{sec:Intro}
The promise that 5G and beyond hold for supporting revolutionary applications (such as autonomous vehicles, intelligent factories, and Internet of Things (IoT)) is contingent on those systems meeting unprecedented performance requirements in terms of achievable rates, latency, and reliability \cite{DLCoordBeam,Rappaport2019,sutton2019enabling}. Communication systems in high-frequency ranges, e.g., millimeter-wave (mmWave) and sub-terahertz (sub THz), present a way to meet the first of those demands. This is primarily due to their abundance of bandwidth that helps achieve data rates in excess of tens of Gbps \cite{What5G?, Rappaport2019}. However, high-frequency systems face challenges on many levels, and one of the most critical challenges is the relatively large beam-training overhead. Signal propagation in the high-frequency domain is characterized by poor penetration ability and suffers from high power loss due to scattering \cite{HeathJr2016}. Therefore these systems need to periodically update the choice of beamforming vectors at both transmitters and receivers to maintain satisfactory Signal-to-Noise Ratios (SNRs). Such need is a common source of strain for those systems, and it has driven much research in the wireless community for innovative solutions to reduce the training overhead.

Some recently emerging approaches to deal with many high-frequency wireless communication challenges revolve around machine learning, and computer vision \cite{ViWi,CamPredBeam,ViWi_blk_pred,serv_id,ViWi-BT,tx_id,charan2021c, imran2023environment}. Those approaches collectively define the \textit{Vision-Aided Wireless Communications} (ViWiComm) framework. Within that framework, a wireless system utilizes computer vision, multimodal machine learning, and deep learning to develop an understanding of the wireless environment and its elements. The system, then, taps into that understanding to address some of the adversities it faces, like the beam-training overhead. A key advantage of the ViWiComm framework is the information-rich sensors it introduces to the wireless system, RGB cameras \cite{charan2021c, charan2022drone}, radars \cite{demirhan2022radar, 10049816}, and LiDAR sensors \cite{jiang2022lidar}, to name a few. These sensors provide much-needed information about the wireless environment that is commonly under-utilized or ignored altogether.

Much of the work on addressing beam training \cite{CamPredBeam,Ying2020} with ViWiComm either assumes simplified wireless settings or is based on synthetic datasets like the ViWi dataset \cite{ViWi}. One may question the practicality of the developed framework, given that real wireless communication environments are characterized by two inherent properties: dynamism and visual diversity. Dynamism in wireless environments refers to the continuous change in the locations of radio transmitters and receivers, which naturally leads to time-varying wireless channels. This property is a serious and well-acknowledged challenge in wireless communications \cite{DetChPred,DLCoordBeam}, for it is the main reason that beam training needs to be performed frequently. Furthermore, dynamism partially contributes to the second property, visual diversity. The wireless environment is fairly visually complex; it is composed of visually diverse objects (e.g., trees, buildings, people, cars, buses, etc.), some of which are continuously on the move, causing the visual scene to vary. From a vision perspective, visual diversity is a serious challenge to a ViWiComm as it leads to the multi-candidate dilemma; it is the case when the composition of the wireless environment includes multiple objects that could constitute a possible wireless transmitter, see \cite{tx_id}.  By considering these two properties, the practicality issue of ViWiComm could be summarized in the form of two questions:
\begin{quote}
	\textbf{Q.1:} \textit{Could the encouraging results obtained with synthetic datasets (e.g., \cite{CamPredBeam,Ying2020}) be extended to datasets collected from real wireless environments?}\\
	\textbf{Q.2:} \textit{Could the vision-aided wireless communication framework tackle the beam-training challenge in wireless environments with multiple candidate transmitters?}
\end{quote} 

Answering these two questions constitutes the cornerstone of this study. More specifically, the study builds on top of the work in \cite{tx_id,charan2021c,DeepSense} and provides a detailed evaluation of the beam training challenge with the ViWiComm framework.

\subsection{Prior Work}
The beam-training challenge in high-frequency wireless systems has been investigated in several studies like \cite{Wang2009,Hur2013}, to name a few. In recent years, a considerable amount of research has explored machine-learning-based approaches to tackle that challenge. For this study, that research is surveyed from the perspective of computer vision. This means the proposed approaches in the literature will be categorized based on whether they utilize visual data or not.

\textbf{Beam prediction without visual data:} Many studies have considered developing machine learning algorithms for beam prediction using wireless and/or position sensory data. Good examples could be found in \cite{DLCoordBeam,Sub6PredMmWave,Wang2019}. The work in \cite{DLCoordBeam} introduces a novel coordinated beamforming approach based on mmWave omni- or quasi-omni-channels. A set of coordinating mmWave base stations estimates their local mmWave channels using omni or quasi-omni beam patterns and then uses those channels to train a Deep Neural Network (DNN) to predict the beamforming vectors at each base station. The study in \cite{Sub6PredMmWave} takes a different look at what could be used as sensory data. It proposes a new approach based on sub-6 GHz channels and DNNs. It poses the beam prediction problem as a classification problem to which a DNN is trained to observe the sub-6 GHz channels and predict the optimal beamforming vector. The use of position sensory data to tackle the beam-training challenge has been explored in \cite{Wang2019}. It proposes to use receiver position and types of neighboring vehicles to handcraft a per receiver feature vector, and it trains an ensemble classifier based on a random forest model to predict the optimal beamforming vectors using that feature vector.

The existing work has shown promise in addressing the beam-training challenge. However, there is potential for further improvement by considering better sensory data and learning methods. Currently, the chosen sensory data is insufficient to capture the complexity and dynamics of the wireless environment. The wireless or position data only offer partial information about objects and their movements. Additionally, the previous research relies solely on unimodal machine learning. Future wireless systems, including base stations and user equipment, are expected to incorporate multiple sensors such as RGB cameras, sub-6 GHz transceivers, LiDARs, radars, and GPS \cite{alkhateeb2023real, jiang2023digital}. Utilizing all available sensory data instead of relying on a single modality would be more practical. Therefore, machine learning research should focus on developing algorithms that can effectively extract meaningful features by combining different modalities.

\textbf{Beam prediction with visual data:} 
This category includes beam prediction approaches that address the two shortcomings mentioned above by utilizing visual sensory data and multimodal machine learning. Using computer vision to address the beam-training challenge in high-frequency wireless systems is rooted in the early work in \cite{ViWi, CamPredBeam}. The ViWiComm framework, in which wireless systems are equipped with visual data sources, has first been introduced in \cite{ViWi} along with the first Vision-Wireless (ViWi) dataset. Using ViWi, \cite{CamPredBeam} presents the first case study for beam prediction using the ViWiComm framework. It proposes a beam prediction approach based on Convolutional Neural Networks (CNNs) for wireless settings with a single-candidate user. The work in \cite{Ying2020} extends that in \cite{CamPredBeam} by utilizing object detectors. It takes a step closer to studying ViWiComm in realistic wireless settings by synthesizing images with multiple candidate users using a scenario from the ViWi dataset. 

Some common limitations of the early work include focusing on simple communication scenarios, such as settings with single-candidate users or artificially generated multi-candidate users, and not fully exploring the potential of multimodal machine learning. Wireless environments, as mentioned earlier, are characterized by dynamic changes and visual variations. This poses a significant challenge when predicting beams based solely on visual data, as visual information alone does not provide any insight into the identity of the object responsible for the radio signal. To address this issue, previous studies such as \cite{tx_id, VAR} have proposed multimodal machine learning algorithms that combine both visual and wireless data to identify the radio transmitter in the environment. However, it is important to note that these studies do not specifically tackle the beam-training challenge in practical wireless settings.

\subsection{Contribution}\label{sec:contributions}
Recognizing both the shortcomings of initial work on ViWiComm and its potential to deal with the beam-training challenge in high-frequency wireless systems, this paper presents a comprehensive study tackling the challenge in practical wireless settings. In particular, it addresses the two questions \textbf{Q.1} and \textbf{Q.2} by utilizing visual data and deep learning models. The main contributions of this paper could be summarized as follows:
\begin{itemize}
	\item \textbf{Beam prediction in single-candidate settings:} As a stepping stone, this paper addresses \textbf{Q.1} using a dataset constructed using DeepSense 6G scenarios \cite{DeepSense}, which represent real wireless communication environments. It shows that those encouraging results in \cite{CamPredBeam,Ying2020} could indeed be achieved in real wireless communication settings with a single-candidate user.
	\item \textbf{Beam prediction in multi-candidate settings:} 
	An answer to \textbf{Q.2} is provided in the form of a novel multimodal beam prediction DNN. The proposed solution utilizes visual and position data to identify the radio transmitter in the environment and predict its optimal beamforming vector. It represents the first demonstration of a multimodal ViWiComm DNN designed to tackle beam prediction in practical wireless communication settings. 
	\item \textbf{Large multimodal dataset and comprehensive evaluation:} 
	By utilizing the different scenarios in DeepSense, two datasets of co-existing multimodal data points (i.e., visual, position, and wireless) are constructed for development and testing purposes. Those datasets constitute the seeds for the comprehensive evaluation experiments conducted to answer the questions in \textbf{Q.1} and \textbf{Q.2} and to provide insights into the upsides and downsides of vision-aided beam prediction. 
\end{itemize} 

\section{Vision-aided Beam Prediction}\label{sec:vabp}
Overcoming the challenge of beam training in high-frequency wireless communication systems (mmWave and sub-THz) is a cornerstone in realizing the full potential of those systems. Recent efforts to deal with this challenge have seen increasing interest in leveraging artificial intelligence, and deep learning in particular \cite{Sub6PredMmWave,DLCoordBeam,CamPredBeam,Ying2020}. Among those efforts, vision-aided beam prediction is a promising approach proposed to reduce the beam training overhead in high-frequency systems. As stated in Section \ref{sec:contributions}, this work provides a comprehensive and realistic evaluation of the potential of vision-aided beam prediction solutions. In the following three subsections, we (i) motivate the need for vision, (ii) introduce the key idea of vision-aided beam prediction, and (iii) present the overall flow of this paper. 

\subsection{Motivation}\label{sec:vabp_motiv}

\begin{figure}[t]
	\centering
	\includegraphics[width=0.8\linewidth]{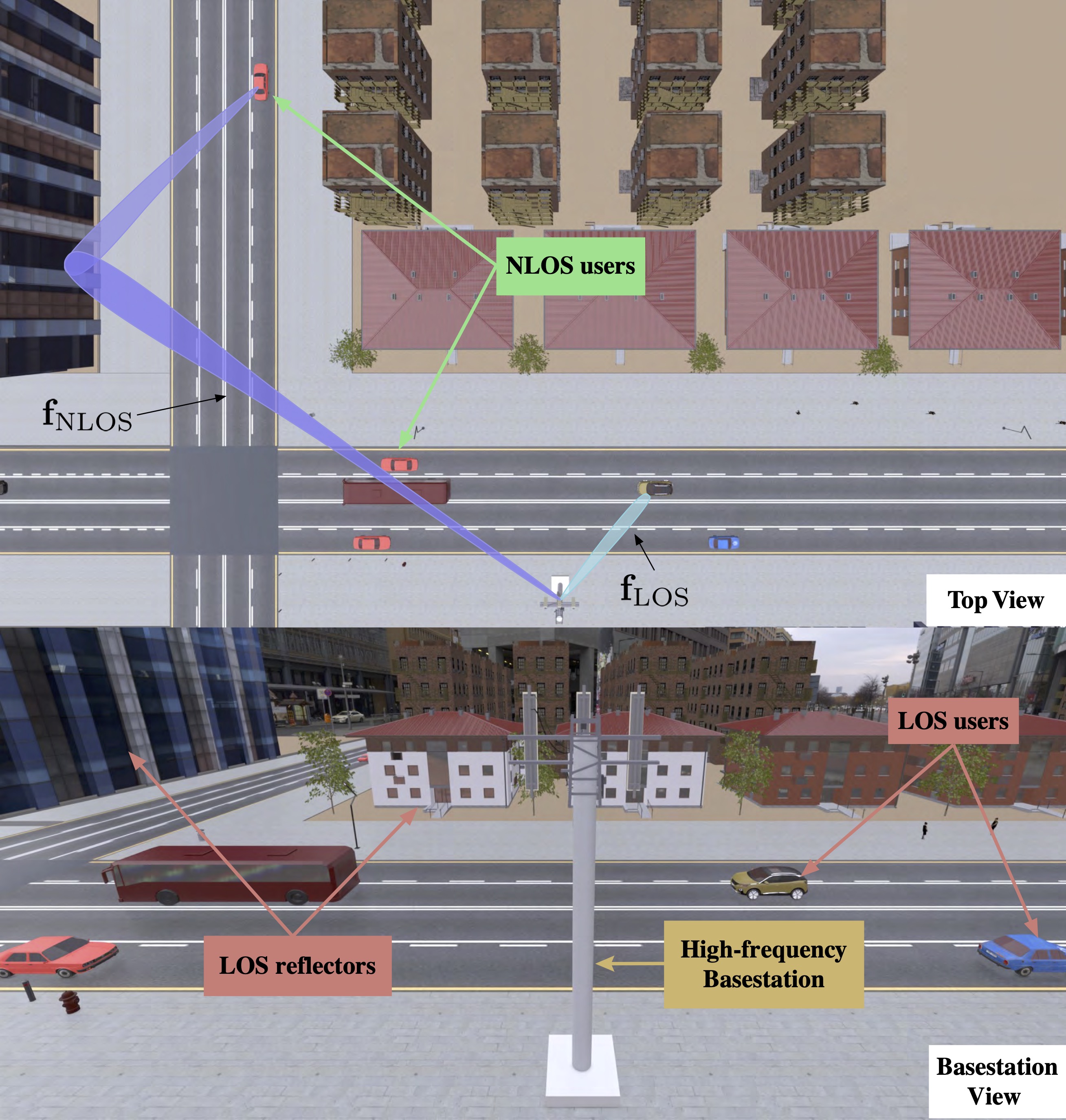}
	\caption{An illustration of a high-frequency wireless communication system and its environment. ``Base station View'' depicts the environment from the base station perspective, showing some LOS users and possible LOS reflectors. ``Top View'' shows the invisible part of that environment, i.e., NLOS users, and the beamforming vectors used to serve LOS and NLOS users.}
	\label{fig:key_idea}
\end{figure}

Harvesting the large bandwidth available at the mmWave and sub-THz bands requires using narrow beams and overcoming their alignment challenge \cite{Rappaport2019,Andrews2016}. The true challenge with beam alignment stems from the inherently dynamic nature of the wireless communication environment; transmitters, receivers, and scatterers could all be on the move. A direct implication of that dynamics is that transmitters and receivers need to periodically update their choice of beamforming vectors to maintain a satisfactory SNR level and, preferably, a LOS connection. This update comes in the form of beam-training, which is a recognized burden in high-frequency wireless communications \cite{Sub6PredMmWave,DLCoordBeam}. 

An interesting and novel approach for handling the beam-training burden could be found in embracing a striking resemblance between high-frequency communication and computer vision systems, which is their reliance on LOS \cite{ViWi_blk_pred,2021GC_BlkPred}. Since high-frequency signals struggle in penetrating objects in the wireless environment and lose a significant amount of power due to scattering \cite{Andrews2016}, there is a quite large SNR margin between LOS and NLOS communication links that skews in favor of LOS. This makes LOS a preferable setting in high-frequency communications, and it draws a connection with computer vision, which is inherently LOS. The data usually captured and analyzed in a computer vision system depicts what is \textit{visible} in the scene, starting with simple patterns (e.g., edges, colors, etc.) to abstract concepts (e.g., human, dog, tree, etc.). As such, the information contained in visual data could be as valuable to a high-frequency system as it is to a computer vision system, \textit{begging the question of how computer vision could be used to mitigate the beam-training challenge}.

\subsection{Key Idea}\label{sec:vabp_key}
The reliance on LOS is the key feature that links high-frequency communications to computer vision and is the bedrock of vision-aided beam prediction. To understand the connection and how it is used to overcome the large beam training overhead challenge, consider the example depicted in \figref{fig:key_idea}, where a high-frequency basestation serves some users in its surrounding environment. Without loss of generality, the base station is assumed to employ a well-calibrated high-frequency antenna array \footnote{Well-calibrated here means the array geometry is known, and no hardware impairments are assumed, more on those issues could be found in \cite{CBLNNets,CBRL}}. Some users experience LOS connections with the base station in this environment, while others have NLOS links. In a classical system operation, the base station performs beam training to identify suitable beams for each user. Two examples are illustrated in the ``top-view'' image in \figref{fig:key_idea}, namely beams $\mathbf f_{\text{LOS}}$ and $\mathbf f_{\text{NLOS}}$. An important factor in the beam choice is the user's relative position with respect to the base station. For instance, $\mathbf f_{\text{LOS}}$ is clearly a LOS beam, and it represents a straightforward path between the base station and the green SUV. The NLOS beam $\mathbf f_{\text{NLOS}}$, does not point to the red car but to a visible building (i.e., reflector) that leads to the red car. Although simple, these observations highlight an important property that could be utilized to address the beam-training challenge; both the SUV and the building are in the field of view of the base station and are qualified as LOS objects. This suggests that detecting those objects and understanding their roles could be key to identifying the choice of beamforming vectors.

A high-frequency wireless system could tap into the advances in the fields of computer vision and machine learning \cite{DLBook} to realize the notion of detecting objects and understand their roles, and perform vision-aided beam prediction. Abstractly, using a machine learning algorithm designed for ViWiComm, the task of predicting beams could be broken down into three stages: \textit{scene analysis}, \textit{object-role identification}, and \textit{decision making}. The first stage is where the machine learning model directly operates on the visual data to extract contextual information. Referring back to \figref{fig:key_idea}, this is loosely equivalent to detecting various objects of importance in the scene, like cars, buses, pedestrians, trees, buildings, etc. Those objects are passed on to the second stage, in which the machine learning attempts to identify the roles of those objects relative to the wireless system; this means labeling the objects in \figref{fig:key_idea} as candidate users, candidate reflectors, or candidate signal blockages. Such labeling will be discussed under the concept of transmitter identification in Section~\ref{sec:multi_cand}. This will highlight that visual data alone is not sufficient for identifying the transmitter in the scene. Hence, the object-role identification stage requires augmenting what has been learned from the previous stage with other sensory data, such as wireless, LiDAR, or GPS data. Recognizing the role of each object gets the machine learning ready to go into the decision-making stage, where it selects the suitable beams to serve the LOS/NLOS users.  

\subsection{Flow}\label{sec:vabp_flow}
This work presents a real-world study on realizing vision-aided beam prediction and demonstrating its ability to tackle the beam-training challenge. The study is conducted in two major phases: (i) Beam-prediction in single-candidate settings, where there is only one candidate user in the visual scene and (ii) beam-prediction in multiple-candidate settings, where multiple objects/candidate users exist in the visual scene. The two phases represent the evolution of the ViWiComm framework first proposed in \cite{CamPredBeam} from conception to real-world implementation in realsitic multi candidate user settings. This study takes advantage of the recently developed DeepSense 6G dataset \cite{DeepSense} that reflect real wireless communication environments. The dataset is designed for multimodal machine learning research in wireless communication. It consists of various scenarios where multimodal sensing and communication data samples are collected using a multi-sensor testbed; see \cite{DeepSense} for more information.

\section{System Model}\label{sec:sys_mod}
This paper adopts a system model that consists of a basestation, deployed on the sidewalk, and a vehicular mobile user, similar to the system depicted in Fig.~\ref{fig:key_idea}. The base station is equipped with a uniform linear array (ULA) with $M$ elements, a standard-resolution RGB camera, and a GPS. For practicality \cite{Sub6PredMmWave}, the base station is assumed to employ analog-only architecture with a single RF chain and $M$ phase shifters. The base station adopts a predefined local beamforming codebook $\boldsymbol{\mathcal F}=\{\mathbf f_q\}_{q=1}^{Q}$, where $\mathbf{f}_q \in \mathbb C^{M\times 1}$ and $Q$ is the total number of beams. This codebook spans a ULA field of view of $\gamma^{\circ}$ along the azimuth plane. Adopting an OFDM with a cyclic prefix of length $D$ and a number of subcarriers $K$, the received downlink signal at the mobile unit is given by
\begin{equation}\label{eq:sys_mod}
	y_{k} = \mathbf h_{k}^T \mathbf f x + \beta_k,
\end{equation}
where $y_{k}\in \mathbb C$ is the received signal at the $k$th subcarrier, $\mathbf f\in \boldsymbol{\mathcal F}$ is the selected beamforming vector, $\mathbf h_{k} \in \mathbb C^{M\times 1}$ is the channel between the BS and the mobile unit at the $k$th subcarrier, $x\in \mathbb C$ is a transmitted complex symbol that satisfies the following constraint $\mathbb E\left[ |x|^2 \right] = P$, where $P$ is a power budge per symbol, and finally $\beta_k$ is a noise sample drawn from a complex Gaussian distribution $\mathcal N_\mathbb C(0,\sigma^2)$. The choice of the beam is determined using classical beam training, in which the base station sweeps the codebook $\mathcal F$ looking for the optimal vector $\mathbf f^{\star}$. Formally, that sweep could be expressed by
\begin{equation}\label{eq:best_beam}
	\mathbf f^{\star} = \underset{\mathbf f_q\in \mathcal F}{\text{argmax}} \frac{1}{K}\sum_{k=1}^{K} \log_2\left( 1 + \mathrm{SNR}|\mathbf h_{k}^T \mathbf f_q |^2 \right),
\end{equation}
where $\mathrm{SNR}$ is the signal-to-noise ratio. However, in LOS-dominated system operation (almost single-path channels), \eqref{eq:best_beam} could be approximated by
\begin{equation}\label{eq:best_beam_approx}
	\mathbf f^{\star} = \underset{\mathbf f_q\in \mathcal F}{\text{argmax}} \frac{1}{K}\sum_{k=1}^{K} |\mathbf h_{k}^T \mathbf f_q |^2.
\end{equation}
The channel model adopted throughout this paper is the geometric mmWave channel model. This model choice comes as a result of two facts: (i) the model captures the limited scattering property of the mmWave band \cite{Alkhateeb2015,DetChPred}, and (ii) the experimental results in this paper are based on real measurements, which are captured well by the geometric model. The channel vector $\mathbf h_k$ in \eqref{eq:sys_mod} is given by
\begin{equation}
	\mathbf{h}_{u,k} = \sum_{d=0}^{D-1} \sum_{\ell=1}^L \alpha_\ell e^{- \j \frac{2 \pi k}{K} d} p\left(dT_\mathrm{S} - \tau_\ell\right) \ba\left(\theta_\ell, \phi_\ell\right),
\end{equation} 
where $L$ is number of channel paths, $\alpha_\ell, \tau_\ell, \theta_\ell, \phi_\ell$ are the path gains (including the path-loss), the delay, the azimuth angle of arrival, and the elevation angle of arrival, respectively, of the $\ell$th channel path. $T_\mathrm{S}$ represents the sampling time while $D$ denotes the cyclic prefix length (assuming that the maximum delay is less than $D T_\mathrm{S}$).


\section{Single-Candidate Settings} \label{sec:single_cand}
This section starts the first phase of this study, where vision-aided beam prediction is studied in a real wireless setting with a single candidate user. These settings are the immediate and natural extension to those in \cite{CamPredBeam}. This phase is structured into two parts. The first part includes the formal definition of the single-candidate beam-prediction problem, the proposed deep learning solution to that problem, and a brief discussion on the practical challenges associated with that solution, motivating the transition to the second phase of this study. The second part of this phase presents a detailed evaluation of the proposed solution on the real-world dataset, and it will be discussed in Section \ref{sec:single_results}.

\begin{figure*}[!t]
	\centering
	\includegraphics[width=0.85\linewidth]{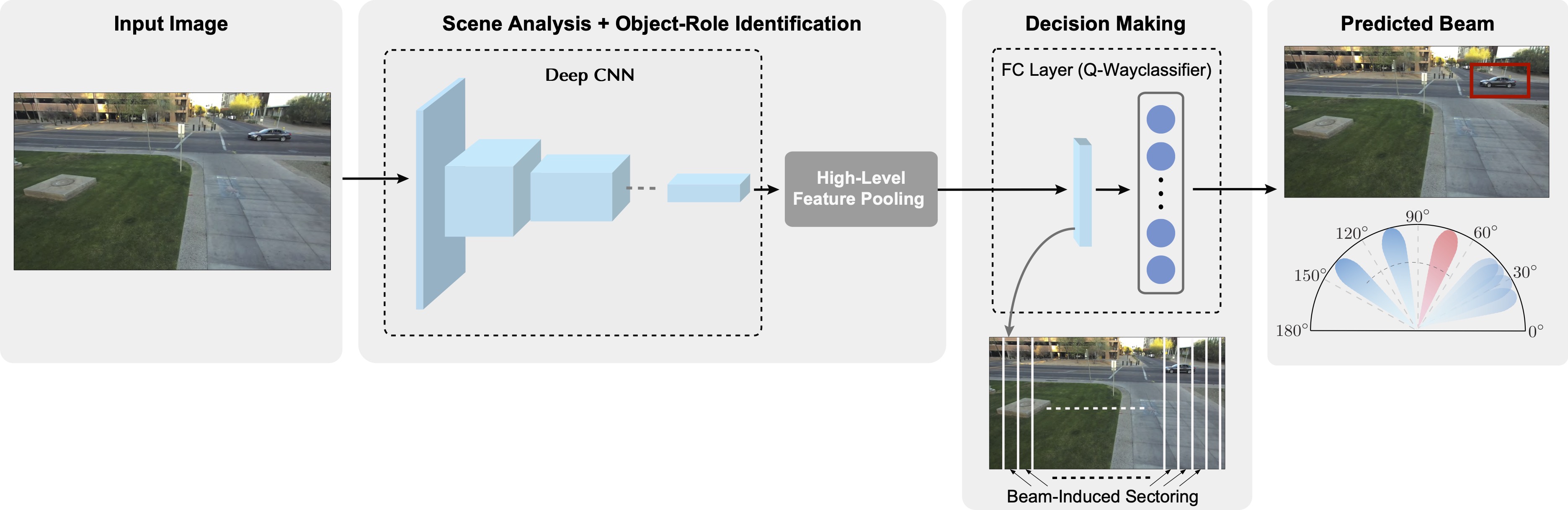}
	\caption{This figure illustrates the proposed machine learning-based vision-aided beam prediction model that leverages visual data captured at the base station for mmWave/sub-THz beam prediction in a single-candidate setting.}
	\label{ref:single-candidate-main-fig}
\end{figure*}

\subsection{Problem definition}\label{sec:single_prob_form}
The beam prediction task in single-candidate wireless settings is defined as follows:
\begin{quote}
	\textit{Given a wireless communication environment where there is only one possible high-frequency transmitter, the vision-aided beam prediction at the infrastructure is the task of predicting the optimal beam indices from a pre-defined codebook by utilizing a machine learning model and the images captured by the camera installed at the basestation}
\end{quote} 
Formally, the problem can be defined as follows. A dataset of image-beam pairs (samples) is collected from real wireless environments where each pair has an image with a single-candidate transmitter and its best beamforming vector. This dataset could be given by $\mathcal D_{\text{task}_1} = \{(\mathbf X_u, \mathbf f^\star_u)\}_{u=1}^U$ where $\mathbf X_u\in\mathbb R^{H\times W\times C}$ is the RGB image of the $u$th pair in $\mathcal D_{\text{task}_1}$ with height $H$, width $W$, and number of channels $C$, and $U$ is the total number of pairs in the dataset. At any time instant $t$, the objective of the single-candidate beam prediction task is to find a prediction/mapping function $f_{\Theta_1}$ that utilizes the available sensory data $\mathbf X[t]$ to predict (estimate) the optimal beam index $ \hat{\mathbf f}[t] \in \boldsymbol{\mathcal F}$ with high fidelity. The mapping function can be formally expressed as
\begin{equation}
	f_{\Theta_1}: \mathbf X[t] \rightarrow  \hat{\mathbf f}[t].
\end{equation}
In this work, we design a deep learning algorithm to learn a prediction function $f_{\Theta_1}$ parameterized by a set of parameters $\Theta_1$ from the dataset $\mathcal D_{\text{task}_1}$. The objective of the learned function is to maximize the overall correct prediction probability over all samples in $\mathcal D_{\text{task}_1}$, which could be expressed as follows
\begin{equation}\label{eq:prob_form_1}
	f^{\star}_{\Theta_1^{\star}} = \underset{f_{\Theta_1}}{\text{max}}\\ \prod_{u=1}^U \mathbb P\left( \hat{\mathbf f}_u = \mathbf f^{\star}_u \ | \ \mathbf X_u \right),
\end{equation}
where the product is the result of an implicit assumption that the samples of $\mathcal D_{\text{task}_1}$ are drawn independently from an unknown distribution $\mathbb P(\mathbf X, \mathbf f^\star)$ that models the relation between $\mathbf X$ and $\mathbf f^\star$\footnote{A common assumption in the realm of machine learning, see \cite{PatternRecog} for example.}.

\subsection{Proposed solution}\label{sec:single_cand_prop_sol}  

The proposed deep learning solution relies on the important parallel between vision and high-frequency communications, i.e., the dependency on LOS objects. It attempts to learn beam prediction by learning where the user (for LOS situations) or reflecting surface (for NLOS situations) are in the environment. In general, a beam-steering codebook deployed at the basestation, especially with a well-calibrated mmWave phased array \cite{CBLNNets}, induces sectoring of the wireless environment across the azimuth plane. This sectoring could be projected onto the image plane, i.e., $\mathbf X$, to result in visual sectors \cite{tx_id} that could be regarded as classes. Given the assumption of a single candidate in the environment, the location of the user (LOS situations) or reflector (NLOS situations) in the image defines the sector to which that user or reflector belongs, henceforth referred to as the object-sector assignment. Therefore, the prediction function $f_{\Theta}(\mathbf X)$ should learn such an assignment in order to predict the best beamforming vector. For the NLOS situations, recognizing the most likely reflector may need a sequence of images; they could help indicate the user direction before it gets blocked and provides some insight into its best reflector.  

Based on the intuition mentioned above about the single-candidate beam prediction task, this study proposes a modified residual neural network (ResNet) \cite{resnet}, more specifically ResNet-50. The architecture is pre-trained on the ImageNet dataset and modified to incorporate a new $M$-class classifier layer. The reason behind the choice is rooted in two main facts. The first one is that residual learning is an effective approach to building very deep architecture. The results in \cite{resnet} have shown that residual blocks, the fundamental element of ResNets, prevent the performance degradation commonly associated with training deep architectures. The second reason is the good performance ResNets have registered in many computer vision tasks. They have been originally designed for image classification; however, they have found their way into many deep architectures developed for object detection \cite{He2017} and semantic segmentation \cite{Dai2016}.

\subsection{Challenges}\label{sec:challenges}

The proposed ResNet architecture above is expected to face a critical challenge when deployed in real mmWave environments. The primary reason is the assumption of a single candidate user, i.e., the proposed solution searches for a single candidate user in the scene to assign to a sector. As a result, the proposed solution is incapable of handling a scenario with multiple users. For instance, when two vehicles are present in the image (the wireless environment), and they belong to different visual sectors, it is impossible to predict the beam without identifying the roles of each vehicle (transmitter, moving blockage, or possible reflector). This dilemma with multiple candidates is the building block for the third phase of this study, which is discussed in the following section.

\section{Multi-Candidate Settings}\label{sec:multi_cand}

The interest in addressing the dilemma of beam prediction in environments with multi-candidate users is at the center of the second phase of this study. As discussed in Section \ref{sec:challenges}, a vision-aided beam prediction algorithm needs, in some way or another, to realize the three-stage process described in Section \ref{sec:vabp_key}. The key challenge in doing so is the ability to identify the roles of every candidate in the environment and differentiate the connected users from the other objects in the environment. In order to deal with this challenge, we need to perform what we call \textit{user identification} in the visual scene by leveraging other user attributed (that could also be captured using other sensing modalities). In this section, we investigate this problem and propose a DNN architecture to perform the task of beam prediction in multi-candidate settings. This establishes the bases for enabling vision-aided multi-user communications. 

\begin{figure*}[!t]
	\centering
	\includegraphics[width=0.9\linewidth]{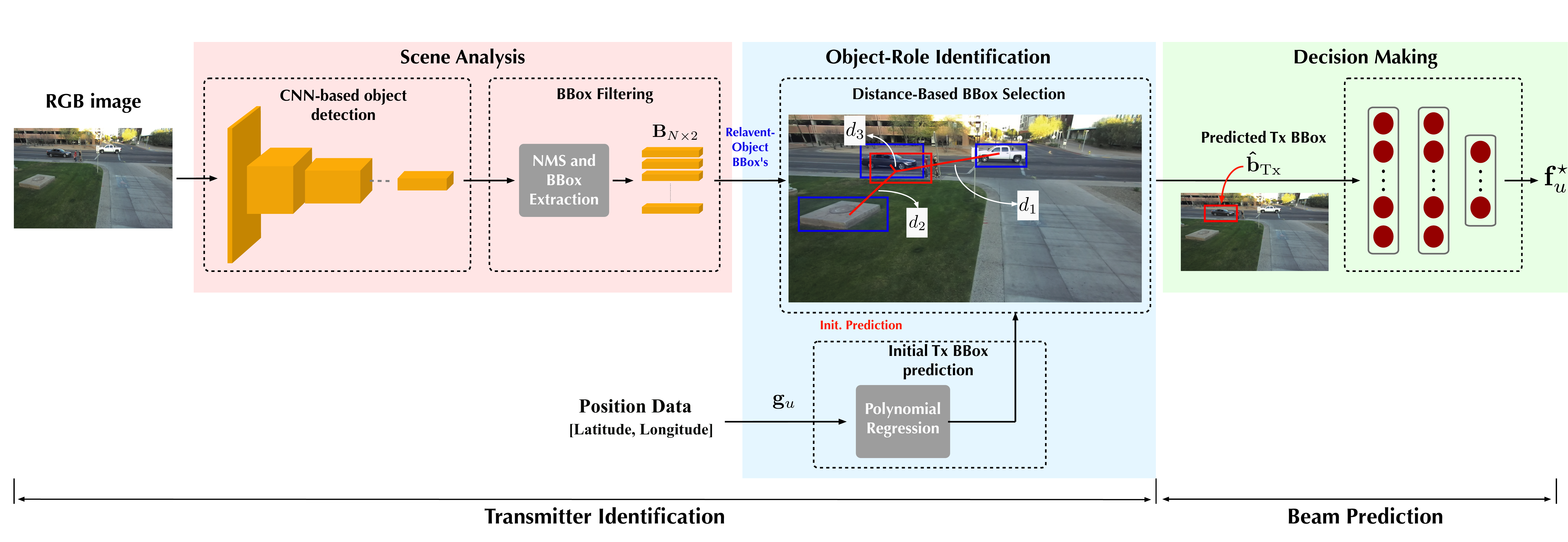}
	\caption{The figure presents the proposed multi-modal machine learning based beam prediction model that leverages both visual and positional data to predict the optimal beam indices in a multi-candidate settings. }
	\label{fig:multi-modal-main-fig}
\end{figure*}

\subsection{Problem Definition}
The beam prediction task in multi-candidate wireless settings is defined as follows:
\begin{quote}
	\textit{Given a wireless communication environment where there are multiple objects that could visually constitute wireless transmitters, the beam prediction task in multi-candidate settings is defined as the problem of predicting the optimal beamforming vector from a pre-defined beam codebook using a pool of multimodal data that includes vision.}
\end{quote} 
The task is formally defined as follows: Let $\mathcal{V}$ be a $v$-tuple of multimodal data samples that includes vision, i.e., $\mathcal V = (\mathbf X, \mathbf g_1, \dots, \mathbf g_v)$ where $\mathbf g_1$ to $\mathbf g_v$ are vectors containing the other modality data samples. Then, a dataset of ($v+1$)-tuples is collected from a real wireless environment $\mathcal D_{\text{task}_2} = \{(\mathcal V_u, \mathbf f^\star_u)\}_{u=1}^U$ where $U$ is the total number of samples in $\mathcal D_{\text{task}_2}$ and $\mathbf f^\star_u$ is the optimal beam in $\boldsymbol{\mathcal F}$ associated with the $u$th $v$-tuple $\mathcal V_u$. At any time instant $t$, the objective of the single-candidate beam prediction task is to find a prediction/mapping function $f_{\Theta_2}$ that utilizes the available sensory data $\mathcal V[t]$ to predict (estimate) the optimal beam index $ \hat{\mathbf f}[t] \in \boldsymbol{\mathcal F}$ with high fidelity. The mapping function can be formally expressed as
\begin{equation}
	f_{\Theta_2}: \mathcal V[t] \rightarrow  \hat{\mathbf f}[t].
\end{equation}
In this work, we design a deep learning algorithm to learn a prediction function $f_{\Theta_2}$ parameterized by $\Theta_2$. This function needs to maximize the probability of correct beam prediction given the image and position data, i.e.,
\begin{equation}\label{eq:prob_form_2}
	f^{\star}_{\Theta_2^{\star}} = \underset{f_{\Theta_2}}{\text{max}}\\ \prod_{u=1}^U \mathbb P\left( \hat{\mathbf f}_u = \mathbf f^{\star}_u \ | \ \mathcal V \right).
\end{equation}
Again, similar to the formulation of \eqref{eq:prob_form_1}, the product in \eqref{eq:prob_form_2} is a result of an implicit assumption that the samples of $\mathcal D_{\text{task}_2}$ are independent and identically distributed, i.e., follow the same unknown joint distribution $\mathbb P(\mathcal V, \mathbf f^\star)$. 
 
\subsection{The Choice of Data Modalities for User Identification}
From the definition above, the ultimate objective of beam prediction in multi-candidate settings is still the same as that in Section \ref{sec:single_prob_form}, developing a deep learning algorithm that predicts optimal beamforming vectors; however, performing the beam-prediction task requires more than visual data in those settings. This is a direct consequence of the fact that multiple candidate users share the same visual traits, see Section \ref{sec:vabp_key} and therefore, their roles from the wireless system perspective \textit{cannot be visually determined}. This calls for a secondary source of information that could augment visual data, and the choice for such source in this study is GPS (i.e., $\mathcal V = (\mathbf X, \mathbf g)$ where $\mathbf g\in\mathbb R^{2}$ is a vector of latitude and longitude coordinates). It is motivated by two important observations. First, position information is intimately related to visual information; GPS provides x-y coordinates for objects in the 3-dimensional world, and an image is a projection of that 3-dimensional world onto a 2-dimensional plane. Hence, in some sense, the position data complements what is missing in the visual data, which is the sense of distance. The second reason is that position data is lightweight, making them easily exchangeable between a basestation and its candidate user. It is quite important to emphasize at this point that GPS (or position data) is \textit{meant to augment visual data and not replace them}. As shown in \cite{charan2021c}, position data could be of help to beam prediction, yet they lack a sense of surrounding; position data only indicate where candidate users are, and they do not account for contextual information (shapes and relations between those candidates). For instance, visual data reflect information about the shape and type of candidates (e.g., large vehicle, small vehicle, pedestrians, etc.) and their relation to one another, which provide contextual information to the machine learning algorithm.

\subsection{Proposed Solution}\label{ref:prop_sol_multi}
This subsection presents the proposed solution for beam prediction in a real-wireless environment with multiple transmitting candidates. It proposes a novel approach that utilizes bimodal visual and position data in $\mathcal D_{\text{task}_2}$ to predict optimal beamforming vectors. The proposed solution follows the three-stage sequence outlined in Section \ref{sec:vabp_key}. More to the point, it breaks down the function $f_{\Theta}(\mathbf X, \mathbf g)$ into two major components: user (transmitter) identification and beam prediction. The first component utilizes visual data to detect relevant objects in the environment (i.e., scene analysis), then identifies the radio transmitter among the objects using position data (i.e., object-role identification). The second component takes in the extracted information about the transmitter and its surrounding objects and predicts the optimal beamforming vector (i.e., decision-making). \figref{fig:multi-modal-main-fig} shows an schematic of the proposed architecture.

\subsubsection{Transmitter Identification} The goal of the first component of the proposed multi-candidate beam prediction solution is to identify the candidate transmitter in the image, i.e., transmitter identification. For that task, a two-step architecture is proposed. The first step of the proposed architecture relies on DNNs to produce bounding boxes enclosing relevant objects in the scene. It is performed to detect all the probable transmitting objects in the environment. In the second step, the DNN uses position data to filter out detected candidates that are not the radio transmitter. A deeper look at the two-step DNN architecture is given below.

\textbf{Bounding box detection:} In order to detect the transmitting candidate in real-wireless settings, the first step is to identify all the relevant objects in the scene (scene analysis). A pre-trained object detector is adopted for this purpose. The object detector is modified to detect two classes of objects in the scene, labeled as ``Tx (transmitter)'' and ``No Tx (Distractors)''. The former label encompasses all objects relevant to the wireless system in the scene. For example, in a scene depicting a city street, relevant objects include, but are not limited to, cars, trucks, buses, pedestrians, and cyclists. The other label includes those cases where no relevant objects are present in the scene. The modified object detector is fine-tuned in a supervised fashion using a subset of the manually labeled dataset described in Section~\ref{sec:dev_data}. In this work, a YOLOv3 architecture is selected for the bounding box detection task. It provides accurate detection at a relatively high frame rate, reducing inference latency. During inference, the fine-tuned YOLOv3 model generates bounding boxes for the detected candidates in the scene and their confidence scores. By using those output boxes, the relevant-object matrix $\textbf{B} \in \mathbb{R}^{N \times 2}$ is constructed such that each row has only the normalized coordinates of the center of a bounding box, see \figref{fig:multi-modal-main-fig}.

\begin{figure*}[!t]
	\centering
	\begin{subfigure}{0.31\textwidth}
		\centering
		\includegraphics[width=1\linewidth]{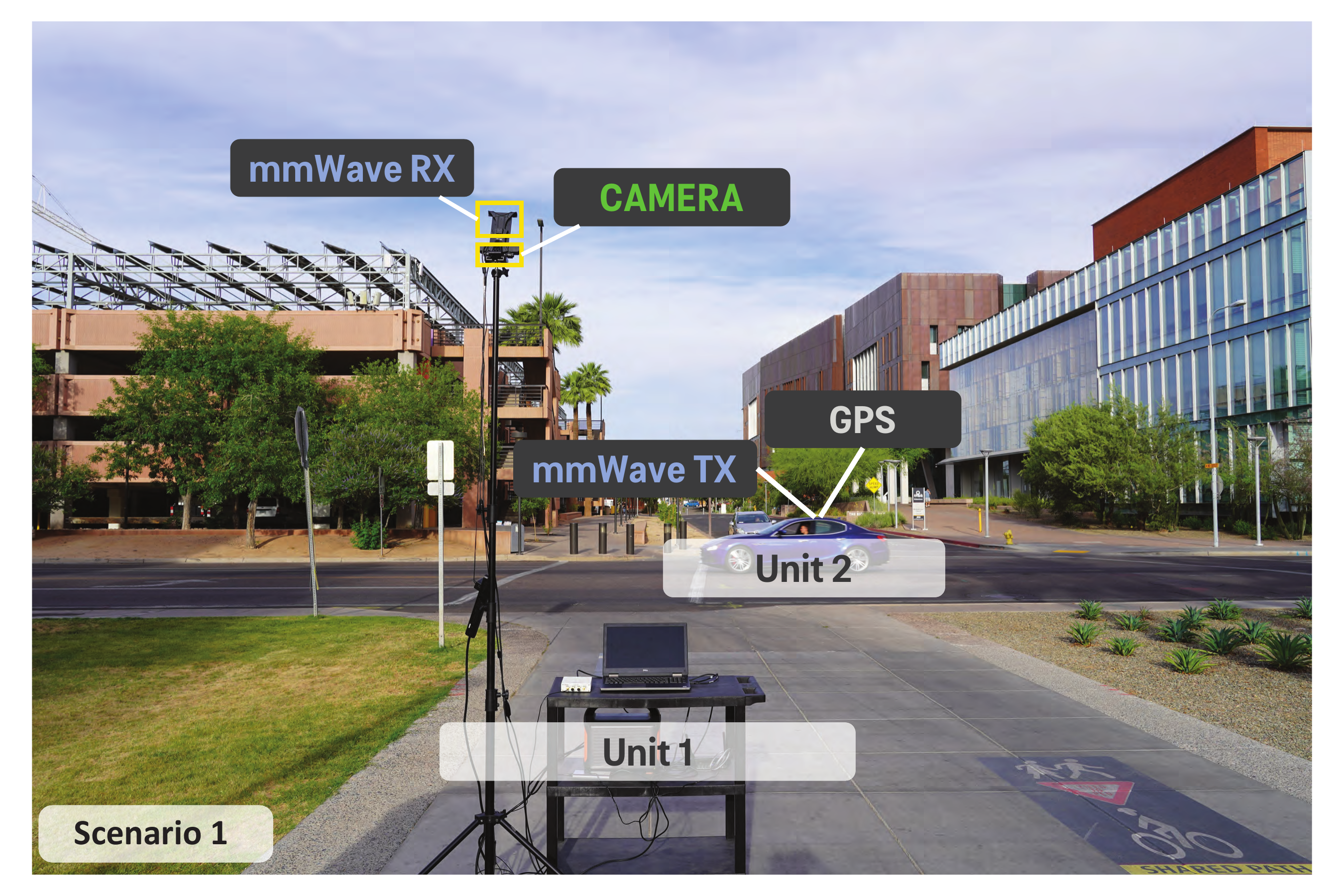}
		\caption{}
		\label{}
	\end{subfigure}
	\begin{subfigure}{0.31\textwidth}
		\centering
		\includegraphics[width=1\linewidth]{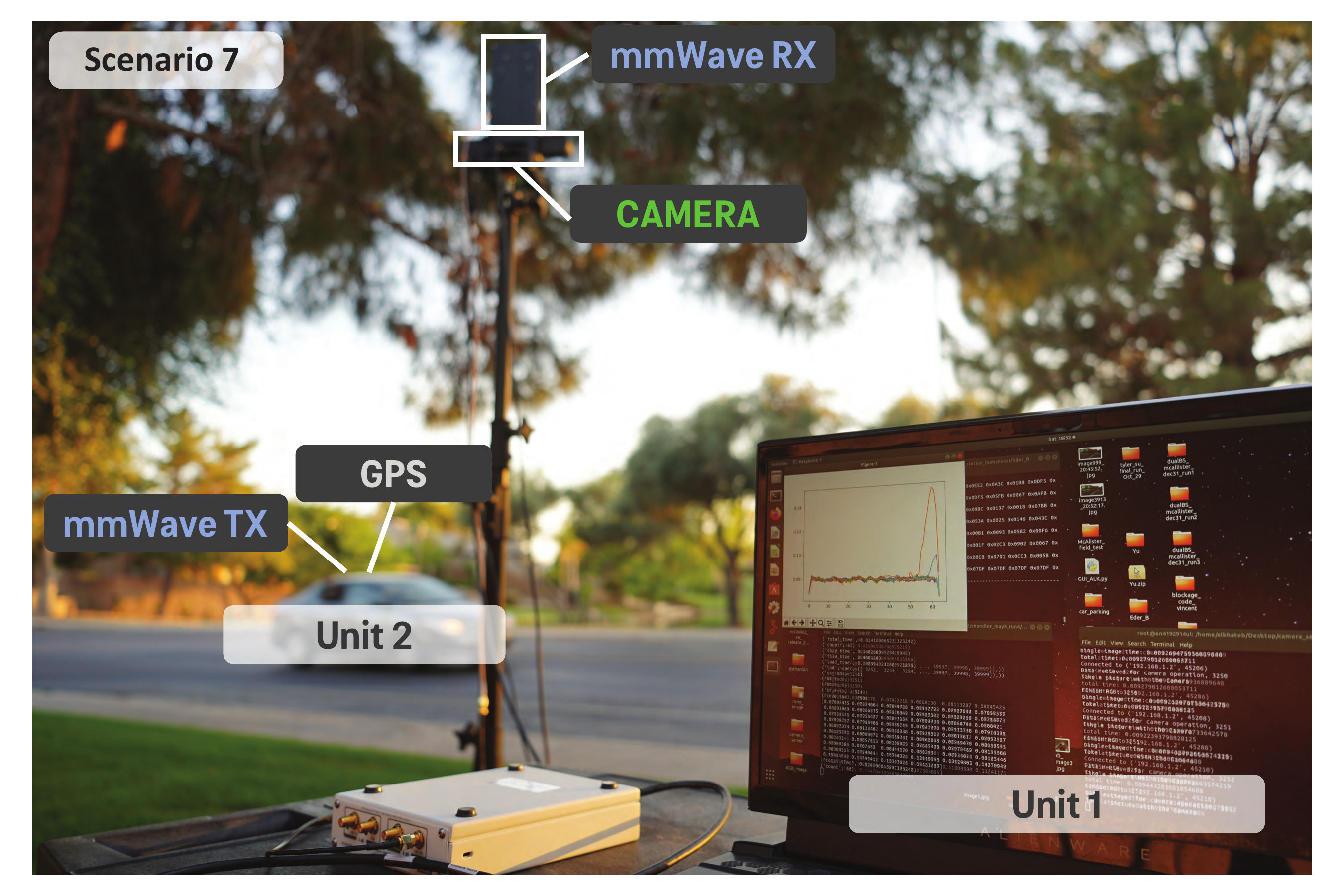}
		\caption{}
		\label{}
	\end{subfigure}
	\begin{subfigure}{0.31\textwidth}
		\centering
		\includegraphics[width=0.975\linewidth]{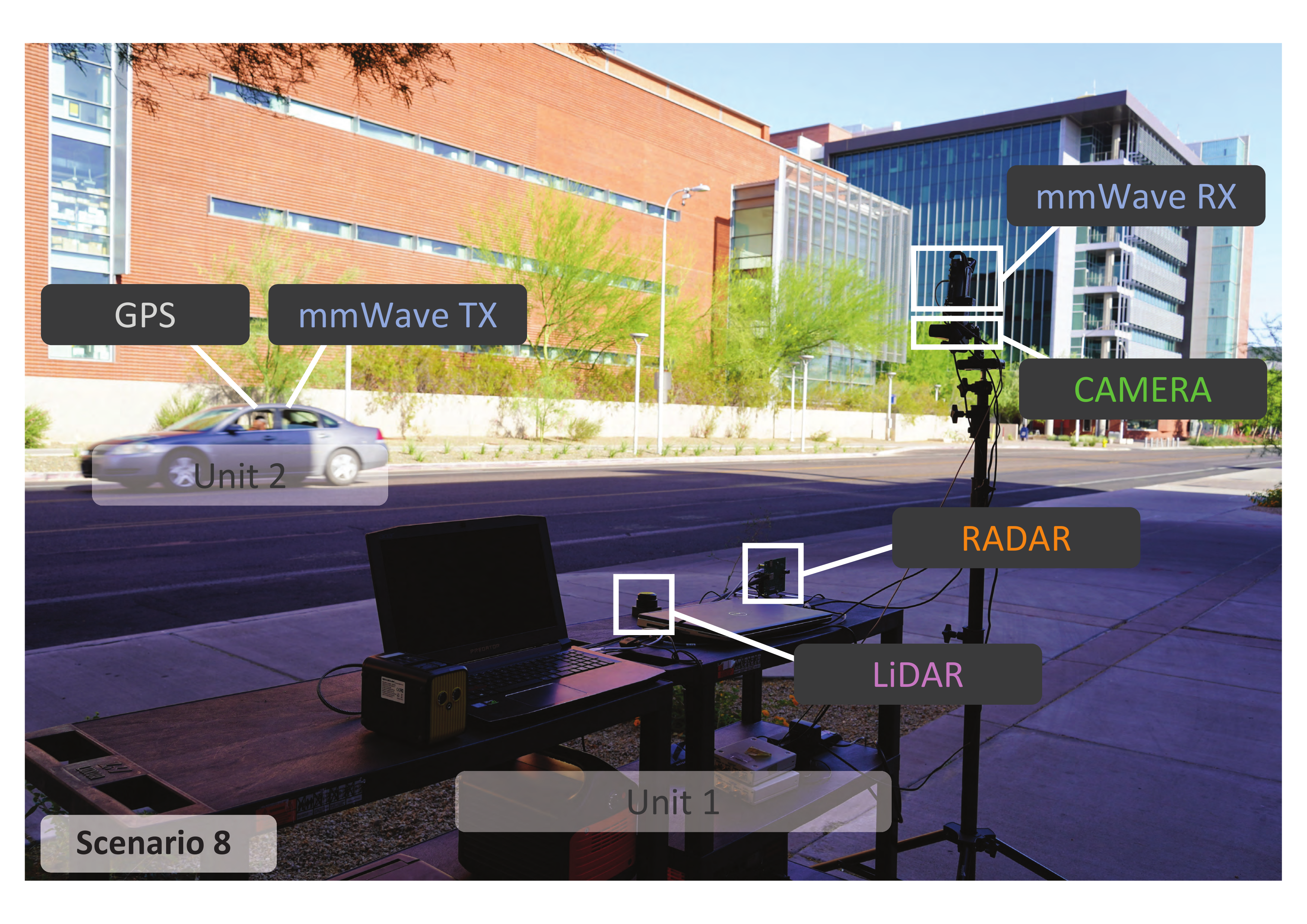}
		\caption{}
		\label{}
	\end{subfigure}
	
	\caption{This figure presents the DeepSense $6$G testbed 1 used during the data collection. It consists of a stationary and mobile unit. The stationary basestation (unit1) is equipped with a mmWave receiver (60 GHz band) and sensory suite (RGB camera, GPS, radar and LiDAR). The mobile unit, acting as a transmitter, is equipped with a 60GHz quasi-omni antenna and GPS receiver.  }
	\label{fig:testbed}
\end{figure*}


\textbf{Bounding box selection:} In this step, both relevant-object matrix $\mathbf B$ and position data are utilized to identify the probable transmitter in the scene. This process starts by learning a prediction function that estimates the bounding box center of a transmitter using its position information, encoding the relation between object position in the 3D world and object location in the image. The function is learned using a $3^{rd}$-degree polynomial regression model
\begin{equation}\label{eq:poly_reg}
	\hat{\mathbf b}_{\text{Tx}} = \mathbf W^T \boldsymbol{\phi},
\end{equation}
where $\hat{\mathbf b}_{\text{Tx}}\in\mathbb R^{2\times 1}$ is a vector with an initial prediction of the centers of a transmitter object, and $\mathbf W$ is an $A\times 2$ parameter matrix, the $A\times 1$ vector $\boldsymbol{\phi}$ is a feature vector obtained from the $3^{rd}$-degree polynomial transformation of the predictors $\mathbf g$, with $A$ denoting the number of unique monomials in a bivariate\footnote{It is bivariate because the vector of predictors $\mathbf g$ is 2 dimensional, see \cite{PatternRecog,multi_var_poly_reg} for more information on polynomial regression.} $3^{rd}$-degree polynomial, i.e., $A = 9$. The parameter matrix $\mathbf W$ is learned from the dataset $\mathcal D_{\text{task}_2}$---more on the training of this model in Section \ref{sec:model_trn}---and once it is learned, the model could be used to get an initial estimate on the center of the transmitter bounding box. Since $\hat{\mathbf b}_{\text{Tx}}$ is an initial estimate that solely relies on position data, it is not expected to be a final prediction but merely a guide. It is used in conjunction with the relevant-object matrix $\mathbf B$ to identify (or select) the object responsible for the radio signal (transmitter). This is done using the nearest neighbor algorithm with an Euclidean distance metric. In other words, the row of $\mathbf B$ with the shortest distance to $\hat{\mathbf b}_{\text{Tx}}$ is picked as the nearest neighbor and, hence, the predicted user (transmitter).

\subsubsection{Beam Prediction} The second component of the proposed solution is a feed-forward neural network that predicts the optimal beam given the identified transmitter. That this component has enough contextual information from the previous one makes it capable of realizing the third and last stage of the three-stage sequence, which is decision-making. A 2-layer feed-forward neural network is developed to perform that beam prediction task. In particular, the prediction task here is posed as a classification problem. The input to the network is the center coordinates of the identified transmitter, and the output is the best beam in the codebook $\boldsymbol{\mathcal F}$.

\section{Development Dataset}
As the cornerstone of this study is to answer \textbf{Q.1} and \textbf{Q.2} in Section \ref{sec:Intro}, an experimental setup is built around multi-modal real-world sensor measurements collected from real wireless environments. This is done by utilizing the publicly available DeepSense dataset \cite{DeepSense} and constructing a large development dataset with tuples of RGB images, mmWave beams, GPS positions, and bounding boxes for the transmitters and distractors. The considered scenarios from DeepSense and the constructed final development datasets are discussed below.


\begin{figure*}
	\centering
	\includegraphics[width=0.8\textwidth]{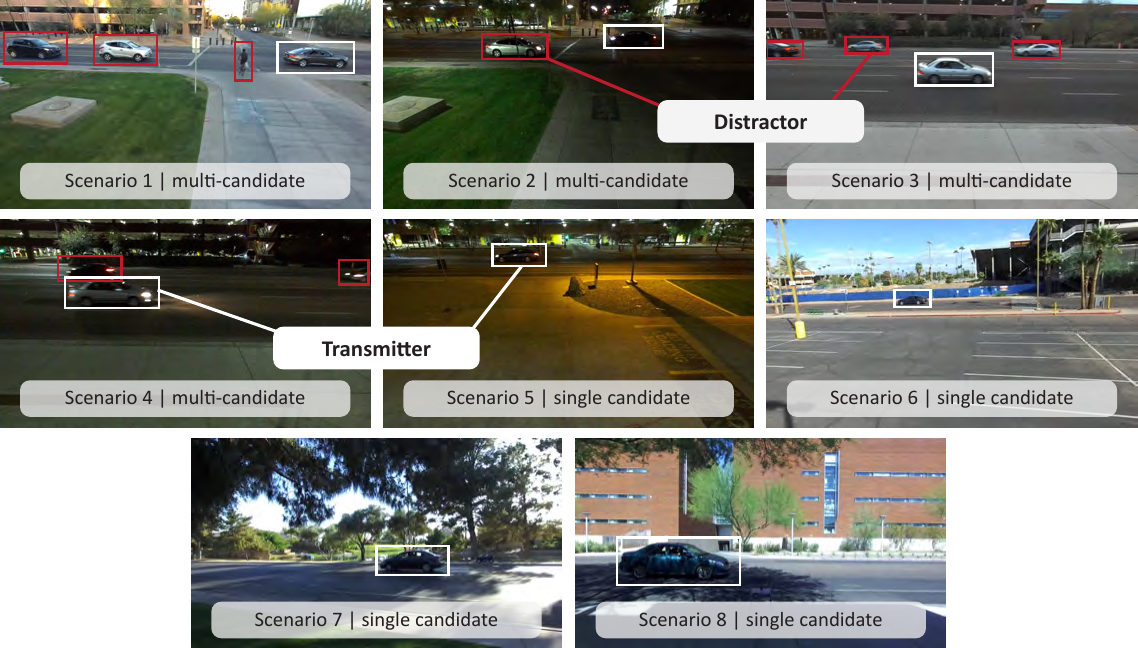}
	\caption{This figure shows the image samples from the different scenarios in the DeepSense $6$G dataset. As shown in this figure, scenarios $1 - 4$ are multi-candidate scenarios, i.e., more than one objects-of-interest (vehicles) are usually present in the FoV of the basestation and have been utilized to investigate the multi-candidate beam prediction problem statement. Scenarios $5-8$ primarily consists of a single object and is useful for evaluating the performance of the proposed sensing-aided single-candidate solution.     }
	\label{fig:img_samples}
\end{figure*}


\subsection{Communication Scenarios and Development Dataset}\label{sec:dev_data}
The DeepSense 6G dataset provides a variety of outdoor wireless communication scenarios with different data modalities \cite{DeepSense}. To evaluate the two-beam prediction problems defined in Sections \ref{sec:single_cand} and \ref{sec:multi_cand}, we select scenarios $1$ to $8$ from the DeepSense dataset. They represent six outdoor wireless environments with vehicles as the main candidate transmitters. Furthermore, the scenarios were collected at different times of the day and in varied weather conditions to increase the overall diversity of the dataset. All eight scenarios contribute a total of $\approx 18000$ data samples, each of which is a tuple of  RGB image, mmWave received power, and GPS position, and across all of them, the wireless system deploys a beam-steering codebook of 64 beams. In Fig.~\ref{fig:img_samples}, we present the data samples from the $8$ different scenarios of the DeepSense dataset. The proposed beam prediction tasks and their solutions mandate slightly different types of communication scenarios and data modalities. Therefore, the raw dataset passes through a processing pipeline to filter out data samples unrelated to a particular task. This leads to two different development datasets, one for each task. The details of this process are given below:

\textbf{Single-candidate dataset:} The task defined in Section \ref{sec:single_cand} requires data collected from the wireless environment with a single candidate in the FoV of the basestation. The first step in generating the development dataset is to filter out samples that do not fit the single-candidate setting from the raw dataset. This is done by manually examining the vision data of each DeepSense scenario. Doing so would show that scenarios 5, 6, 7, and 8 have some multimodal data that pertain to the single-candidate setting. Table \ref{table:tab_single_dataset} lists those scenarios and the number of samples they contribute to the development dataset. The selected samples undergo a second processing step, in which only visual and wireless data is retained. The task relies on visual data as inputs and optimal beams as targets (labels); therefore, only RGB images and mmWave received power data are picked from the modalities of each scenario. The power data go through one extra step to generate the optimal beams. Each received power vector is first downsampled to 32 elements by selecting every other element in the vector. Since the basestation receives the mmWave signal using an oversampled codebook of $64$ pre-defined beams, the downsampling does not affect the total area covered by the beams. This implies that the effective size of the codebook in this paper is $Q= 32$. Then, out of the 32 elements per vector, the index of the beam with maximum received power is selected as the optimal beam (as described in Section \ref{sec:sys_mod} and given by \eqref{eq:best_beam_approx}). The final outcome of this pipeline is a dataset $\mathcal D_{\text{task}_1}=\{(\mathbf X, \mathbf f^{\star})_u\}_{u=1}^{U_1}$ where $U_1\approx 9000$. 

\begin{table*}[!t]
	\caption{Single-Candidate and Multi-Candidate Dataset}
	\centering
	\setlength{\tabcolsep}{5pt}
	\renewcommand{\arraystretch}{1.2}
	\begin{tabular}{|c|l|c|c|c|}
		\hline
		\multirow{2}{*}{\textbf{Dataset}}                               & \multicolumn{1}{c|}{\multirow{2}{*}{\textbf{Location}}} & \multirow{2}{*}{\textbf{Time of Day}} & \multicolumn{2}{c|}{\textbf{Number of Samples}} \\ \cline{4-5} 
		& \multicolumn{1}{c|}{}                                   &                                       & \textbf{Training}     & \textbf{Validation}     \\ \hline \hline
		\multirow{4}{*}{\textbf{Single-Candidate}}                      & \textbf{Tyler Parking (Scenario 5)}                                 & Night                                   &  1817                 & 736                  \\ \cline{2-5} 
		& \textbf{Lot-59 Parking (Scenario 6)}                                & Day                                 &    812               & 348                     \\ \cline{2-5} 
		& \textbf{Downtown Chandler (Scenario 7)}                                & Day                                   &  641                & 275                     \\ \cline{2-5} 
		& \textbf{Bio-design (Scenario 8)}                                      & Day                                 & 3077                   & 1320                       \\ \hline
		\multicolumn{1}{|l|}{\multirow{4}{*}{\textbf{Multi-Candidate}}} & 
		\textbf{McAllister Ave. (Scenario 1)}                                & Day                                   & 1904                  & 816   \\ \cline{2-5}
		\multicolumn{1}{|l|}{}                                          & 
		\textbf{McAllister Ave. (Scenario 2)}                                & Night                                 & 2197                  & 942                     \\ \cline{2-5}
		\multicolumn{1}{|l|}{}                                          & 
		\textbf{Rural Rd. (Scenario 3)}                                      & Day                                   & 1120                  & 477                     \\ \cline{2-5}                    
		\multicolumn{1}{|l|}{}                                          & 
		\textbf{Rural Rd. (Scenario 4)}                                      & Night                                 & 1363                  & 592                     \\ \hline
	\end{tabular}
	\label{table:tab_single_dataset}
\end{table*}

\textbf{Multi-candidate dataset:} The multi-candidate task requires data samples collected from a wireless environment with multiple candidates. Thus, similar to the single-candidate dataset, the raw dataset is examined to identify the multi-candidate samples. This reveals that Scenarios 1, 2, 3, and 4 are more suited for multi-candidate settings, see Table \ref{table:tab_single_dataset} for more details. As described in Section \ref{sec:multi_cand}, the development dataset for this task requires the preparation of data tuples of RGB images, GPS position, and mmWave optimal beams. This means the extra step from the single-candidate dataset is also applied here to obtain the optimal beam indices for each data sample, i.e., downsampling and using \eqref{eq:best_beam_approx}. The final outcome of this simple pipeline is a dataset $\mathcal D_{\text{task}_2}=\{(\mathcal V_u, \mathbf f^{\star})_u\}_{u=1}^{U_2}$ where $U_2\approx  9500$.

The two development datasets above are used to train and evaluate the performance of the proposed solutions. Both datasets are further divided into training and validation sets with a split of $70-30\%$. The details of the experimental setup are provided in the next section, while Section \ref{sec:exp_results} is devoted to discussing the performance of the proposed solutions.

\section{Experimental Setup} \label{sec:model_trn}

This section first presents an overview of the model training process and the hyper-parameters utilized to train the proposed machine learning model for both single-candidate and multi-candidate settings. Next, we discuss the metric used to evaluate the beam prediction performance of the proposed solution. All the experiments are performed on a single NVIDIA Quadro RTX 6000 GPU using the PyTorch deep learning framework.

\textbf{Network Training:} In Section~\ref{sec:single_cand_prop_sol} and Section~\ref{ref:prop_sol_multi}, we proposed two different machine learning-based models for the single-candidate and the multi-candidate settings, respectively. Both the proposed solutions utilize the cross-entropy loss with Adam optimizer to train the models. The detailed hyper-parameters used to fine-tune the models are presented in Table~\ref{tab:nn_train_params}. Next, we present the in-depth details of the model training.

\textbf{Single-Candidate:} As described in Section~\ref{sec:single_cand_prop_sol}, the proposed vision-aided beam prediction solution adopts an ImageNet pre-trained ResNet-50 object classification model. The model is further modified by removing the last output layer and replacing it with a fully-connected layer with $M=32$ neurons. The proposed model is trained in a supervised manner with a dataset $\mathcal D_{\text{task}_1}$, comprising RGB images and its corresponding ground-truth beam index.
	
\textbf{Multi-Candidate:} The multi-candidate proposed solution consists of two major components, namely, (i) transmitter identification and (ii) beam prediction. The transmitter identification component consists of a pre-trained YOLOv3 model, which is further fine-tuned in a supervised fashion using a subset of manually labeled dataset $\mathcal D_{\text{task}_2}$. During inference, the YOLOv3 model is used to detect all the relevant objects and extract the bounding boxes of those objects. The proposed solution utilizes the user's positional data to select the most probable bounding box. After identifying the probable transmitter, the bounding box center coordinates, $\hat{\mathbf b}_{\text{Tx}}$, are then provided as input to the proposed feed-forward neural network. The proposed machine learning model is then trained in a supervised fashion using the ground-truth bounding-box coordinate and the corresponding beam index.

\textbf{Evaluation Metric:} Here, we present the details of the metric adopted to evaluate the efficacy of the proposed solution. The primary metric adopted for evaluation is the top-$k$ accuracy. The top-$k$ accuracy is defined as the percentage of the test samples where the ground-truth beam is within the top-$k$ predicted beams. In this work, we utilize the top-1, top-2, top-3, and top-5 accuracies to compute the prediction performance of the proposed solution. In Section~\ref{sec:exp_results}, we present the in-depth evaluation of the proposed solution for both single-candidate and multi-candidate settings. 

\begin{table}[!t]
	\caption{Design and Training Hyper-parameters}
	\centering
	\setlength{\tabcolsep}{5pt}
	\renewcommand{\arraystretch}{1.2}
	\begin{tabular}{@{}l|cc@{}}
		\toprule
		\toprule
		\textbf{Parameters}                     & \textbf{ResNet-50}  & \textbf{MLP}        \\ \midrule \midrule
		\textbf{Batch Size}                     & 32                  & 32                  \\
		\textbf{Learning Rate}                  & $1 \times 10 ^{-4}$ & $1 \times 10 ^{-2}$ \\
		\textbf{Learning Rate Decay}            & epochs 4 and 8      & epochs 20 and 40    \\
		\textbf{Learning Rate Reduction Factor} & 0.1                 & 0.1                 \\
		\textbf{Dropout}                        & 0.3                 & 0.3                 \\
		\textbf{Total Training Epochs}          & 15                  & 50                  \\
		\textbf{Number of Output Nodes ($M$)}   & 32                  & 32                  \\
		\bottomrule \bottomrule
	\end{tabular}
	\label{tab:nn_train_params}
\end{table}


\section{Experimental Results}\label{sec:exp_results}

The performance of the proposed ViWiComm beam prediction solutions is studied in this section. The discussion is divided into two parts. The first will discuss the performance of the proposed DNN in single-candidate settings. It will highlight the main advantages and shortcomings of ViWiComm for beam prediction, setting the stage for the discussion on the multi-candidate setting. The second part will focus on the multi-candidate settings and show how the proposed solution can handle the beam prediction task in those practical settings.

\subsection{Single-candidate}\label{sec:single_results}

The beam prediction performance of the proposed solution is studied from two different standpoints, machine learning and wireless communication. The first perspective includes experiments that evaluate the performance of the proposed DNN architecture from different machine-learning angles, e.g., performance per location, number of data samples needed for training, etc. Those experiments use machine learning metrics such as top-k accuracy, where $\text{k}\in\{1,2,3,5\}$ is the accuracy rank, and confusion matrices. The second perspective attempts to translate the results of the machine learning evaluation into results pertaining to the wireless system performance, such as studying the implications of beam-prediction failure on the wireless receive power.

\subsubsection{Machine Learning Perspective}\label{sec:ml_persp_sing}
The proposed DNN architecture is evaluated on the single-candidate dataset. This evaluation investigates three important questions: (i) Does the proposed DNN perform consistently across different wireless environments? (ii) Are there any advantages to training the proposed DNN on several environments simultaneously? Furthermore, (iii) how many data samples are needed to learn the beam prediction in single-candidate settings, in general?

\begin{figure*}[!t]
	\centering
	\includegraphics[width=0.65\linewidth]{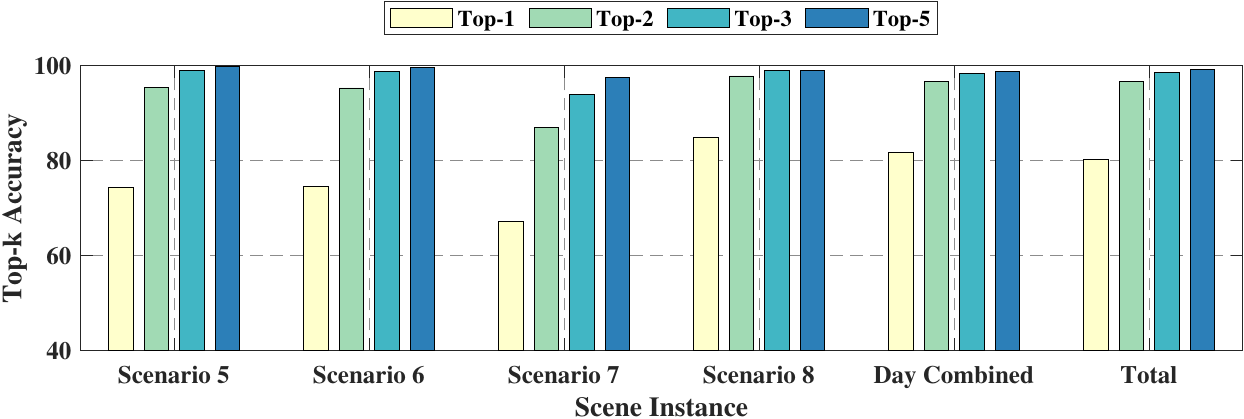}
	\caption{Top-k accuracies ($\text{k}\in\{1,2,3,5\}$) of the proposed DNN across all four single-candidate scenarios and two choices of combining, combining day time scenarios and combining all four scenarios.}
	\label{fig:single_cand_acc}
\end{figure*}

\textbf{Could the DNN have consistent performance across different environments (scenarios)?} This question attempts to identify whether the proposed DNN can achieve similar single-candidate beam-prediction performance across different wireless environments or not. \figref{fig:single_cand_acc} addresses the question by training and testing the DNN on each scenario individually and two combinations of scenarios. The first thing one might observe from the figure is how volatile the top-$1$ performance looks across different environments, ranging from $\approx67\%$ to $\approx84\%$. This could be attributed to the difference between the wireless environments. The four scenarios represent four different physical locations and two different times of day, i.e., scenario $5$ represents a wireless environment operating at night. The fluctuation might initially seem alarming, for it suggests a sense of uneven DNN performance. However, a closer look at the top-$3$ and $5$ performances negates that suggestion; for \textit{it shows that the DNN could produce consistent or nearly uniform performance}. From the figure, the DNN registers almost the same accuracies across locations and times of day, ranging from $\approx92\%$ to $\approx99\%$ for top-$3$---the implications of this will be further explored in Section \ref{sec:wire_persp_sing}.

\begin{figure}[!t]\captionsetup[subfigure]{font=footnotesize}
	\centering
	\begin{subfigure}{0.49\textwidth}
	\centering
	\includegraphics[width=0.65\linewidth]{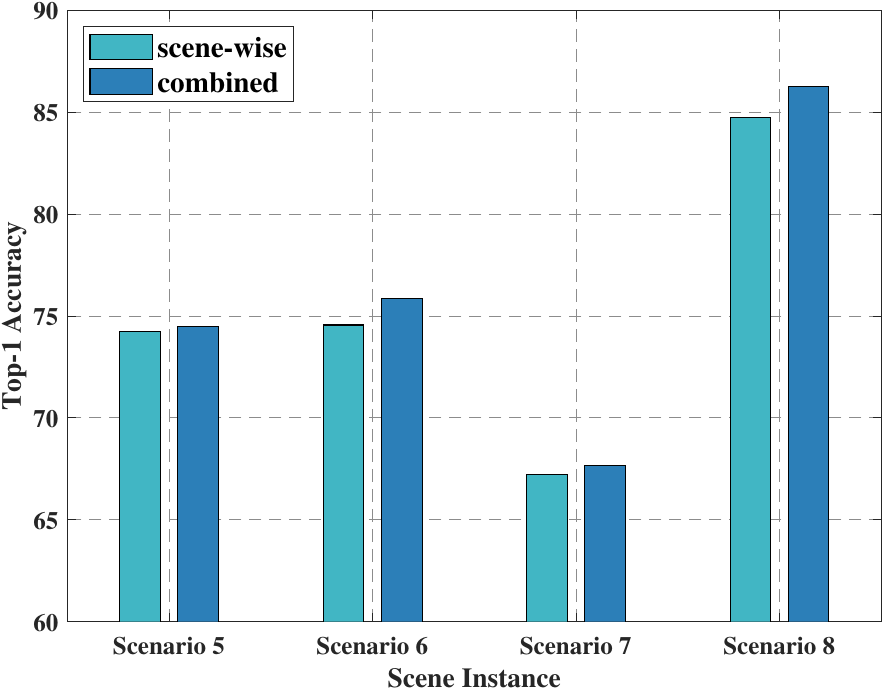}
	\caption{Individual versus combined}
	\label{fig:single_cand_acc_comb}
	\end{subfigure}
	\begin{subfigure}{0.49\textwidth}
	\centering
	\includegraphics[width=0.66\linewidth]{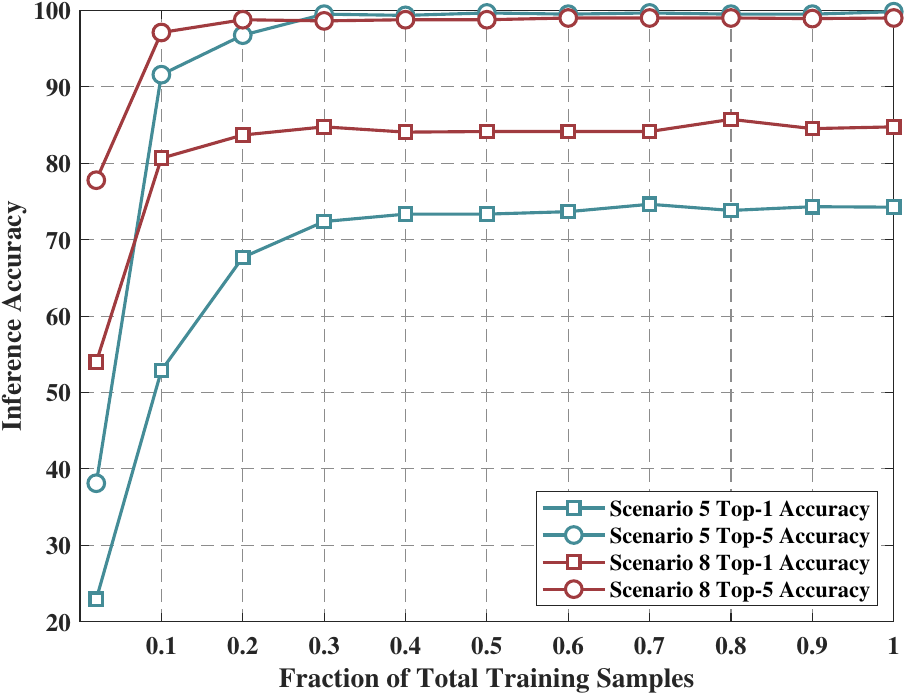}
	\caption{Accuracy versus percentage of samples used}
	\label{fig:single_cand_number_of_samples_comb}
	\end{subfigure}
	\caption{A deeper dive into the performance of the proposed DNN. (a) highlights the value of combining data samples from different scenarios. (b) sheds some light on how many training samples are required to get the performances in \figref{fig:single_cand_acc}.}
\end{figure}

\textbf{What is the gain of learning from multiple scenarios?} Even more interesting than the consistency of the DNN performance is the impact of combining scenarios on that performance. The bars of ``Day Combined'' and ``Total'' in \figref{fig:single_cand_acc} indicate two intertwined facts: (i) A model can be shared across environments and (ii) combining data samples could help improve the top-1 performance on each scenario. When the DNN is trained on a combined dataset (whether combining scenarios having the same time of day or combining all), it achieves a top-$1$ performance that is \textit{better} than the top-1 performance of three of the individual scenarios. It is observed that in the ``Total'' case, the top-$1$ accuracy is higher than those of individual scenarios, i.e., scenarios $5$, $6$, and $7$. \textit{This not only indicates that a single model could be trained for multiple scenarios simultaneously but that combining scenarios can even help in improving the learning process}. At first glance, one could be tempted to attribute this improvement to the unbalanced data contribution of each scenario, see Table \ref{table:tab_single_dataset}; scenario $8$ with its $1320$ validation samples may bias the top-1 performance of the combined dataset. However, this is not the case, and the improvement is a result of the improved learning process for the DNN. \figref{fig:single_cand_acc_comb} corroborates this conclusion. It compares the top-1 accuracy of the DNN when it is trained on each scenario individually and on all four scenarios combined, i.e., training on individual scenarios separately and testing on individual scenarios or training on all scenarios together and testing on individual scenarios.

\textbf{How many training samples are needed?} This interesting question could be seen as a natural follow-up to the previous discussion, for it ponders the computational cost of that performance. \figref{fig:single_cand_number_of_samples_comb} provides an answer to that question. It plots the top-1 and top-5 accuracies versus the number of training samples used for two scenarios. An obvious observation from the figure is that approximately $30\%$ of the total training samples is enough to achieve the reported top-1 performances in \figref{fig:single_cand_acc} and even less than $30\%$ is needed for top-5. Under the surface of this observation lies a more interesting takeaway. The figure consolidates the earlier conclusion on the role of combining scenarios in achieving improved performance. Adding more data samples (more than the $30\%$) does not have much of an impact on the performance of the DNN. This means the improvement observed after combining the scenarios is actually a consequence of an improved learning process and not an increased number of data samples.

\subsubsection{Wireless Communication Perspective}\label{sec:wire_persp_sing}
\begin{figure}[!t] \captionsetup[subfigure]{font=footnotesize}
	\centering
	\begin{subfigure}{0.49\textwidth}
	\centering
	\includegraphics[width=0.65\linewidth]{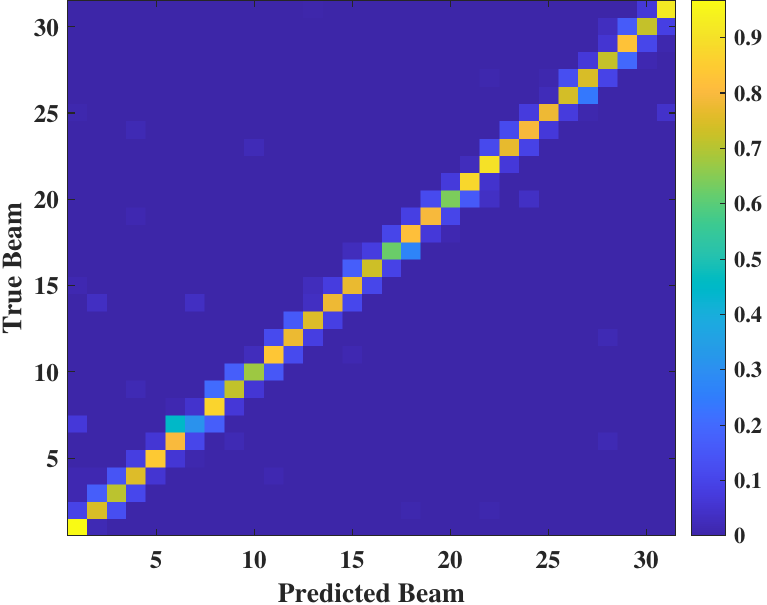}
	\caption{Confusion matrix}
	\label{fig:single_cand_conf_mtx_comb}
	\end{subfigure}
	\begin{subfigure}{0.49\textwidth}
	\centering
	\includegraphics[width=0.66\linewidth]{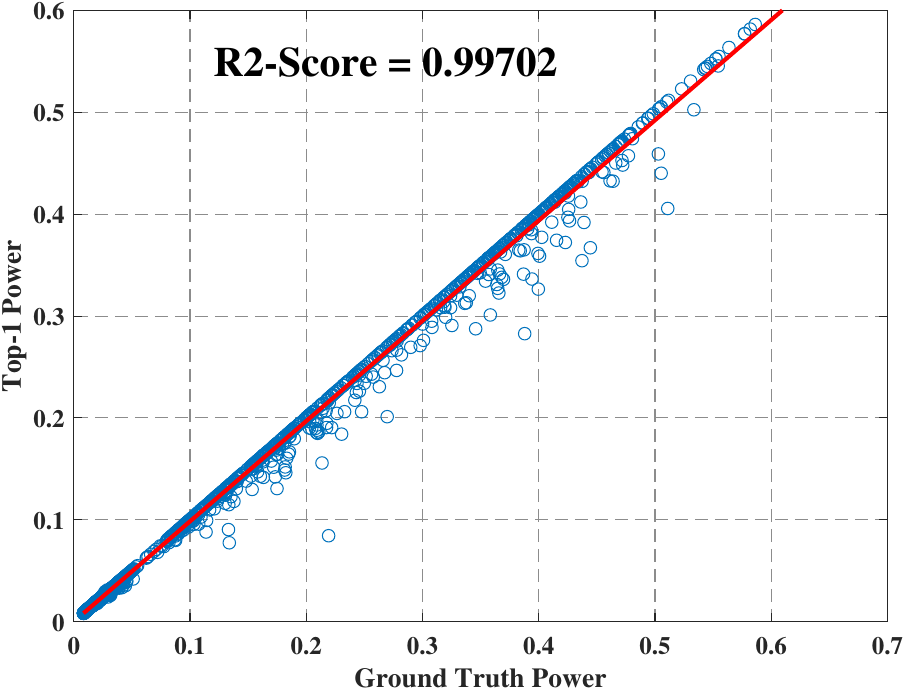}
	\caption{Power scatter plot}
	\label{fig:sing_cand_r_score}
	\end{subfigure}
	\caption{(a) shows the confusion matrix of beam prediction. (b) illustrates the relation between the groundtruth received power and the received power using the top-1 predicted beam.}
	\label{fig:wire_pers}
\end{figure}

This perspective focuses on the implications of the performance of the proposed DNN on the wireless system. More specifically, it attempts to answer two critical questions: (i) What are the implications of predicting the wrong beam? Moreover, (ii) how much of an impact does mis-prediction have on the wireless system?

\textbf{What are the implications of mis-predictions?} The previous results in \figref{fig:single_cand_acc} indicate that the prediction of the proposed solution deviates from the optimal beam between $\approx16\%$ to $\approx33\%$ of the times (based on top-1 accuracy). This may seem concerning at first, yet, as discussed earlier, a closer look at top-3 or 5 shows a much slimmer margin of error, $\approx8\%$ to $\approx1\%$. This means that a wireless system may not need to rely exclusively on the proposed DNN, obsoleting classical beam training. Instead, \textit{the system could use the top-3 or 5 beams in conjunction with some lightweight beam training}; it trains the wireless user on the predicted top-3 or 5 beams to determine the optimal one, which adds a level of robustness to the system operation.

\textbf{What if only the top-1 prediction is used?} This is an interesting question as it encourages a dive into the effect of mis-prediction on the wireless system. \figref{fig:wire_pers} is one way to address that question. Its subfigures are obtained on the case of training and testing the DNN on all scenarios at the same time (case of ``Total'' in \figref{fig:single_cand_acc}). The confusion matrix, \figref{fig:single_cand_conf_mtx_comb}, suggests that even when the DNN misses, its top-1 prediction is likely to be one of the neighboring beams, e.g., if the optimal beam is 15, the DNN is most likely to predict beams between 14 and 16. Below the surface, such mis-prediction is not costly; neighboring beams are expected to achieve reasonable wireless performance. This is verified by \figref{fig:sing_cand_r_score}. It shows a scatter plot for the top-1 received power versus the ground-truth received power. The figure indicates that the neighborhood of the optimal beam registers a similar level of received power. Hence, predicting a neighboring beam achieves almost the same received power as the optimal beam. This is quantified on the figure using the $\text{R}^2$-score, which reflects the compactness of received power compared to the ground-truth.
\begin{figure*}
	\centering
	\includegraphics[width=0.7\linewidth]{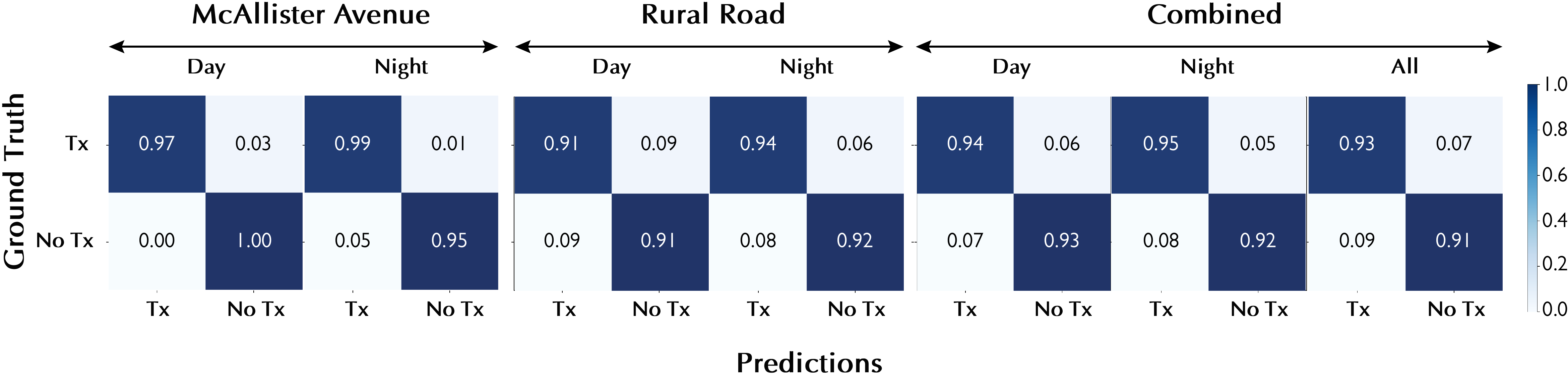}
	\caption{The confusion matrices on all case studies in the multi-candidate settings. Each matrix quantifies the likelihood of identifying a transmitter in a group of candidates with similar visual traits.}
	\label{fig:tx_id}
\end{figure*}

\begin{figure*}
	\centering
	\includegraphics[width=0.65\linewidth]{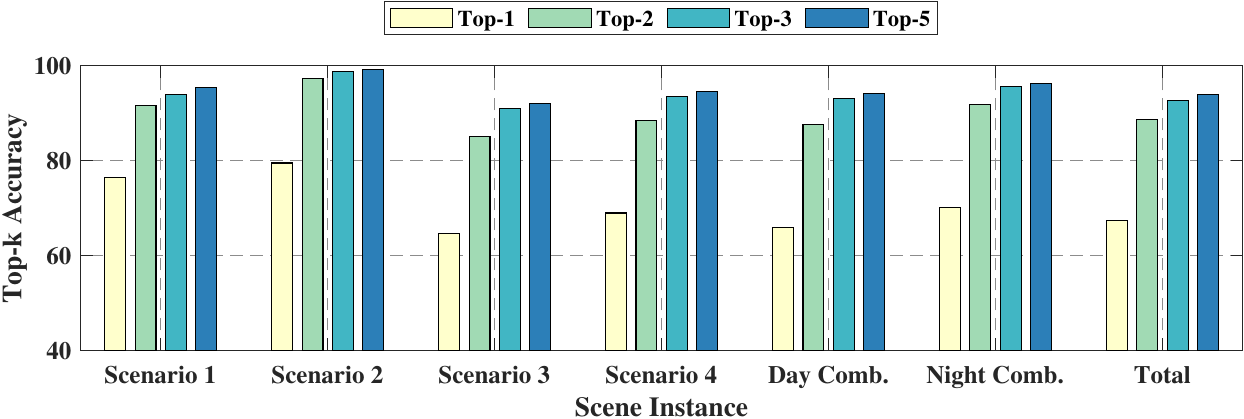}
	\caption{Top-k accuracies ($\text{k}\in\{1,2,3,5\}$) of the proposed DNN across all four multi-candidate scenarios and three choices of combining, combining day time scenarios, night time scenarios, and combining all four.}
	\label{fig:acc_multi}
\end{figure*}

\subsection{Multi-Candidate Beam Prediction}

This section focuses on the practical case of beam prediction in multi-candidate settings. As mentioned in Section~\ref{ref:prop_sol_multi}, the proposed solution has two components, transmitter identification, and beam prediction. Hence, the performance evaluation here is divided into three subsections. The first one evaluates the performance of the transmitter identification component, and the other two evaluate the beam prediction performance in a similar way to that presented in Section \ref{sec:single_results}, namely from machine learning and wireless communication perspectives.

\subsubsection{Transmitter Identification}

As described in Section \ref{ref:prop_sol_multi}, this component performs scene analysis and object-role identification. In particular, it attempts to identify the target user in the visual scene (from the other objects/distractors). The performance of transmitter identification is evaluated on the multi-candidate dataset before the component is integrated with beam prediction. \figref{fig:tx_id} shows seven confusion matrices for transmitter identification, four for individual scenarios and three for different combined scenarios. Overall, the matrices display high true positive and negative rates, which do not go below $\approx 90\%$. This indicates a good precision-recall performance regardless of the location or time of day a scenario represents. For instance, when all four scenarios are combined (case of ``All'' in the figure), the proposed DNN achieves a precision of $\approx91\%$ at a recall of $93\%$. Such good precision-recall performance translates to high confidence in the proposed DNN to identify transmitters in various wireless environments and lighting conditions.

\subsubsection{Machine Learning Perspective}\label{sec:ml_persp_multi}
The two-component DNN is evaluated on the multi-candidate dataset, and the focus of this evaluation is still the same as that of Section \ref{sec:ml_persp_sing}.

\textbf{Could the DNN have consistent performance across scenarios?} Addressing this question is even more interesting in multi-candidate settings than it is in single-candidate settings, as they are the epitome of practical wireless environments. \figref{fig:acc_multi} shows the top-$k$ beam prediction accuracies over all four scenarios and three different combinations of them. The first thing one could observe there is a slight performance degradation across all variants of the top-$k$ metric. The primary reason behind the reduction in model performance is \textit{the multi-candidate nature of these scenarios} (i.e., various relevant objects appear in the RGB image). It makes the beam-prediction task much more challenging than single-candidate. Identifying the user (transmitter), in these cases, is no longer straightforward, as evident from the confusion matrices in \figref{fig:tx_id}. However, a closer look at the top-$3$ and $5$ accuracies highlights the following. Compared to the single-candidate settings, top-3 accuracies fluctuate within a range of $10\%$ (a $3\%$ increase over that in single-candidate), and top-5 accuracies register a range of $\approx5\%$ (a $4\%$ increase over that of single-candidate). \textit{These numbers, overall, are quite encouraging because they indicate a fairly consistent performance considering the increased challenge in multi-candidate settings}.
 
 \begin{figure}\captionsetup[subfigure]{font=footnotesize}
 	\centering
 	\begin{subfigure}{0.49\textwidth}
 		\centering
 		\includegraphics[width=0.65\linewidth]{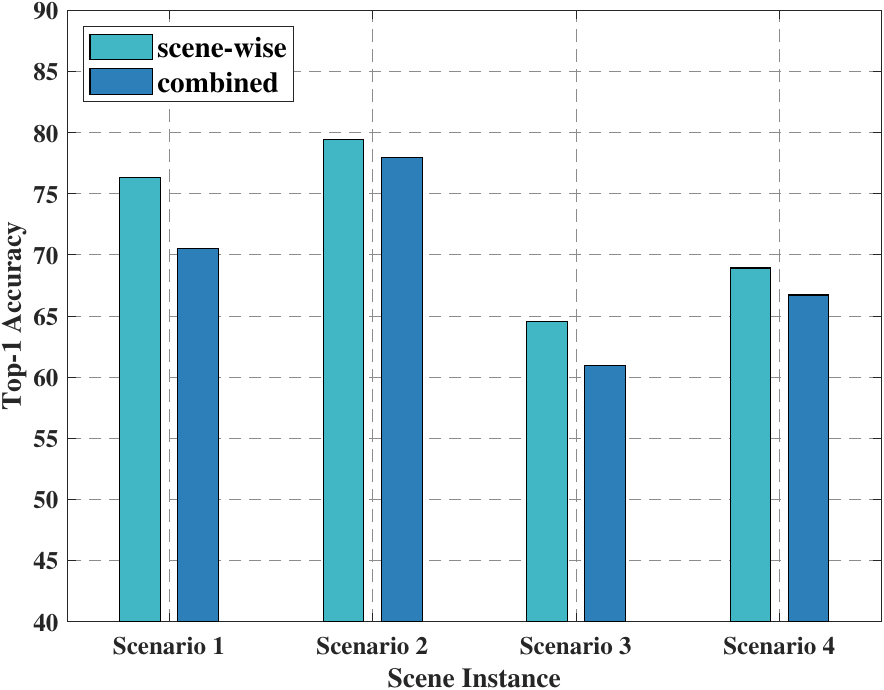}
 		\caption{Individual versus combined}
 		\label{fig:multi_cand_indiv_comb}
 	\end{subfigure}
 	\begin{subfigure}{0.49\textwidth}
 		\centering
 		\includegraphics[width=0.66\linewidth]{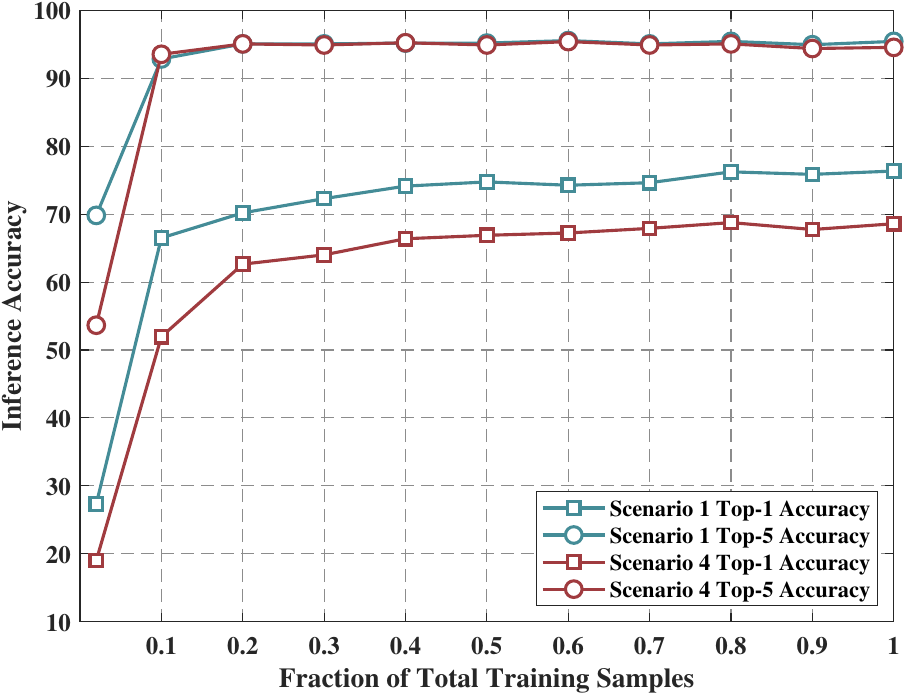}
 		\caption{Accuracy versus percentage of samples used}
 		\label{fig:multi_cand_acc_comb}
 	\end{subfigure}
 	\caption{A deeper dive into the performance of the proposed DNN. (a) explores the impact of combining data samples from different scenarios. (b) sheds some light on how many training samples are required to get the performances in \figref{fig:single_cand_acc}.}
 \end{figure}

\textbf{What is the gain of learning from multiple scenarios?} Combining scenarios and training the proposed DNN does not result in the same improvement in performance seen in the single-candidate case. In fact, combining seems to yield the same or slightly degraded performance as opposed to individual scenarios, as shown in the bars ``Day Comb.'', ``Night Comb.'', and ``Total'' of \figref{fig:acc_multi}. The reason for this could be traced back to the bottleneck of the proposed DNN, which is the transmitter identification component. Owing to its reliance on position data that are inherently noisy\footnote{GPS data do not guarantee zero positioning error. They provide latitude and longitude information within a certain error range.} and the presence of multiple candidate transmitters in an image, a transmitter might not be correctly identified. Despite how rarely this happens, when it does, it leads to significant mis-prediction. This clarifies that mis-identifying the transmitter often leads to critical performance loss as the predicted beam is not in the neighborhood of the optimal one).

\textbf{How many training samples are needed?} The previous discussion has extended the findings from single-candidate beam prediction to multi-candidate cases. More specifically, the proposed DNN displays consistent performance across different scenarios, and the model can be trained on multiple scenarios simultaneously. In parallel to the single-candidate discussion, it is important to establish how many training samples are required to get those results. \figref{fig:multi_cand_acc_comb} presents two examples for training on scenarios 1 and 4. The top-1 accuracies for both examples require approximately $50\%$ of the training samples in each scenario to achieve the reported performance in \figref{fig:acc_multi}. Compared with the need for $30\%$ of the samples in the single-candidate settings, both examples highlight the difficulty of the beam prediction task in multi-candidate settings. 

\subsubsection{Wireless Communication perspective}\label{sec:wire_pers_multi}

\begin{figure} \captionsetup[subfigure]{font=footnotesize}
	\centering
	\begin{subfigure}{0.49\textwidth}
	\centering
	\includegraphics[width=0.65\linewidth]{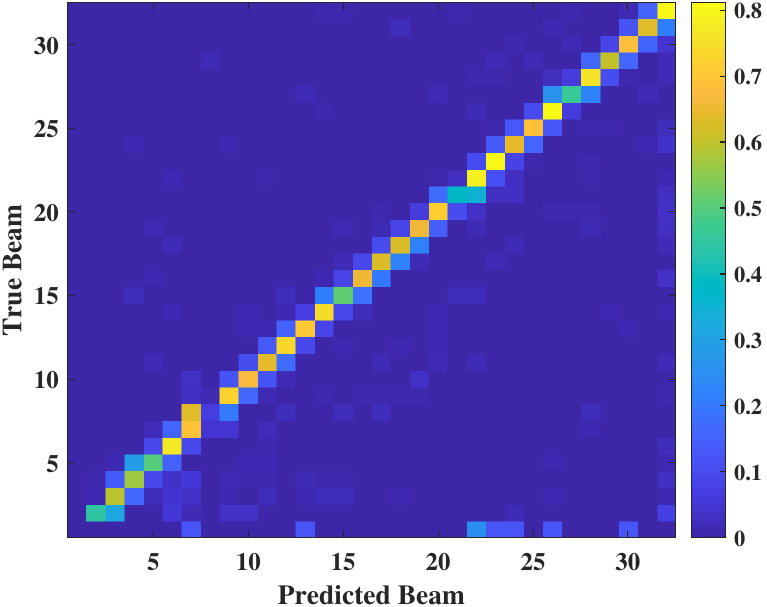}
	\caption{Confusion matrix}
	\label{fig:conf_mtx_multi}
	\end{subfigure}
	\begin{subfigure}{0.49\textwidth}
	\centering
	\includegraphics[width=0.66\linewidth]{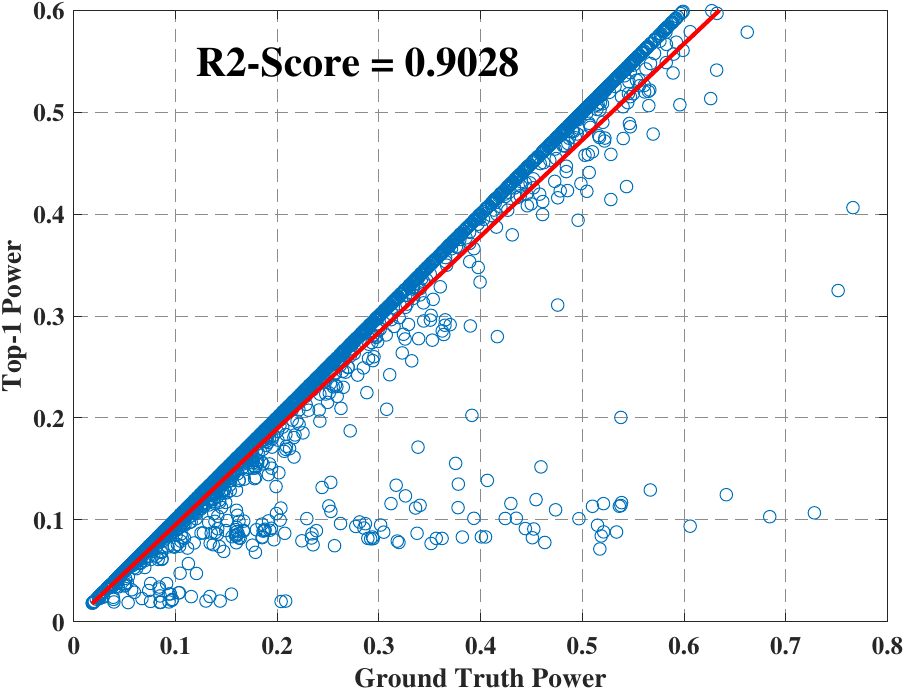}
	\caption{Power scatter plot}
	\label{fig:multi_cand_power_dist_comb}
	\end{subfigure}
	\caption{(a) shows the confusion matrix of beam prediction. (b) illustrates the relation between the groundtruth received power and the received power using the top-1 predicted beam.}
\end{figure}
Next, we investigate the performance from wireless perspective and draw some insights about the system operation. 

\textbf{What are the implications of mis-predictions?}
The findings in Section \ref{sec:ml_persp_multi} indicate a slight dip in the beam prediction accuracies, especially top-1 accuracy. The implications of that on the wireless system performance could be explored further with the confusion matrix in \figref{fig:conf_mtx_multi}. The first thing to observe from the figure is that most predictions identify the optimal beam or a beam in its neighborhood. \textit{This means that robust beam prediction could still be achieved with lightweight beam training and the top-5 predictions of the DNN}, as in the single-candidate settings. However, in contrast to the confusion matrix in \figref{fig:single_cand_conf_mtx_comb}, there is a higher likelihood of seeing predictions far from the ground truth. This is evident from the number of bright spots scattered around the diagonal of the matrix. They are a direct consequence of the catastrophic mis-predictions resulting from misidentifying the transmitter.

\textbf{What if only the top-1 prediction is used?}
The impact of mis-predictions in multi-candidate settings becomes clear when a wireless system relies only on top-1 predictions. \figref{fig:acc_multi} and \figref{fig:conf_mtx_multi} have hinted to the phenomenon, but neither has explored its direct impact on the wireless system performance, and as such, \figref{fig:multi_cand_power_dist_comb} attempts to bridge that gap and round out this analysis. It shows a power scatter plot similar to that presented in \figref{fig:sing_cand_r_score}. The plot shows a wider cluster of blue points, resulting in a smaller $\text{R}^2$-score compared to the single-candidate settings. This score directly reflects the impact of catastrophic mis-prediction; the phenomenon produces beam predictions with very low received power. For instance, the lower half of the y-axis (Top-1 Power) shows some examples where the received power of a top-1 predicted beam is $\approx 0.1$. In contrast, the ground truth beam produces received power in the range of $0.3$ to $0.7$, i.e., ground truth power is $3$ to $7$ times higher than the predicted beam. \textit{Such observation emphasizes the importance of augmenting a ViWiComm system with lightweight beam training in practical wireless environments}.

\section{Conclusion}
This paper presents a machine learning framework specifically designed to address the challenges of realistic scenarios in highly mobile mmWave/sub-THz wireless communication systems with multiple probable transmitting candidates. The proposed solution utilizes visual and positional data to predict optimal beam indices, effectively reducing the beam training overhead associated with adjusting narrow beams of large antenna arrays. Experimental evaluation on the DeepSense $6$G dataset demonstrates that the proposed solution can achieve close to $100\%$ top-5 beam prediction accuracy for single-user scenarios and approximately $95\%$ for multi-object scenarios while accurately identifying the probable transmitting candidate with over $93\%$ accuracy. An important area for future work is ensuring the generalizability of the trained model to unseen scenarios, further enhancing its applicability in real-world mmWave/sub-THz wireless communication networks.

\balance


\begin{thebibliography}{10}
	\providecommand{\url}[1]{#1}
	\csname url@samestyle\endcsname
	\providecommand{\newblock}{\relax}
	\providecommand{\bibinfo}[2]{#2}
	\providecommand{\BIBentrySTDinterwordspacing}{\spaceskip=0pt\relax}
	\providecommand{\BIBentryALTinterwordstretchfactor}{4}
	\providecommand{\BIBentryALTinterwordspacing}{\spaceskip=\fontdimen2\font plus
		\BIBentryALTinterwordstretchfactor\fontdimen3\font minus
		\fontdimen4\font\relax}
	\providecommand{\BIBforeignlanguage}[2]{{%
			\expandafter\ifx\csname l@#1\endcsname\relax
			\typeout{** WARNING: IEEEtran.bst: No hyphenation pattern has been}%
			\typeout{** loaded for the language `#1'. Using the pattern for}%
			\typeout{** the default language instead.}%
			\else
			\language=\csname l@#1\endcsname
			\fi
			#2}}
	\providecommand{\BIBdecl}{\relax}
	\BIBdecl
	
	\bibitem{DLCoordBeam}
	A.~Alkhateeb, S.~Alex, P.~Varkey, Y.~Li, Q.~Qu, and D.~Tujkovic, ``Deep
	learning coordinated beamforming for highly-mobile millimeter wave systems,''
	\emph{IEEE Access}, vol.~6, pp. 37\,328--37\,348, 2018.
	
	\bibitem{Rappaport2019}
	T.~S. {Rappaport}, Y.~{Xing}, O.~{Kanhere}, S.~{Ju}, A.~{Madanayake},
	S.~{Mandal}, A.~{Alkhateeb}, and G.~C. {Trichopoulos}, ``Wireless
	communications and applications above 100 {GHz}: Opportunities and challenges
	for {6G} and beyond,'' \emph{IEEE Access}, vol.~7, pp. 78\,729--78\,757,
	2019.
	
	\bibitem{sutton2019enabling}
	G.~J. Sutton, J.~Zeng, R.~P. Liu, W.~Ni, D.~N. Nguyen, B.~A. Jayawickrama,
	X.~Huang, M.~Abolhasan, Z.~Zhang, E.~Dutkiewicz \emph{et~al.}, ``Enabling
	technologies for ultra-reliable and low latency communications: From {PHY}
	and {MAC} layer perspectives,'' \emph{IEEE Communications Surveys \&
		Tutorials}, vol.~21, no.~3, pp. 2488--2524, 2019.
	
	\bibitem{What5G?}
	J.~G. Andrews, S.~Buzzi, W.~Choi, S.~V. Hanly, A.~Lozano, A.~C. Soong, and
	J.~C. Zhang, ``What will {5G} be?'' \emph{IEEE Journal on selected areas in
		communications}, vol.~32, no.~6, pp. 1065--1082, 2014.
	
	\bibitem{HeathJr2016}
	R.~W. Heath, N.~Gonzalez-Prelcic, S.~Rangan, W.~Roh, and A.~M. Sayeed, ``An
	overview of signal processing techniques for millimeter wave mimo systems,''
	\emph{IEEE journal of selected topics in signal processing}, vol.~10, no.~3,
	pp. 436--453, 2016.
	
	\bibitem{ViWi}
	M.~Alrabeiah, A.~Hredzak, Z.~Liu, and A.~Alkhateeb, ``Vi{W}i: A deep learning
	dataset framework for vision-aided wireless communications,'' in \emph{2020
		IEEE 91st Vehicular Technology Conference (VTC2020-Spring)}, 2020, pp. 1--5.
	
	\bibitem{CamPredBeam}
	M.~Alrabeiah, A.~Hredzak, and A.~Alkhateeb, ``Millimeter wave base stations
	with cameras: Vision-aided beam and blockage prediction,'' in \emph{2020 IEEE
		91st Vehicular Technology Conference (VTC2020-Spring)}, 2020, pp. 1--5.
	
	\bibitem{ViWi_blk_pred}
	G.~Charan, M.~Alrabeiah, and A.~Alkhateeb, ``Vision-aided {6G} wireless
	communications: Blockage prediction and proactive handoff,'' \emph{IEEE
		Transactions on Vehicular Technology}, vol.~70, no.~10, pp. 10\,193--10\,208,
	2021.
	
	\bibitem{serv_id}
	M.~Alrabeiah, U.~Demirhan, A.~Hredzak, and A.~Alkhateeb, ``Vision aided {URLL}
	communications: Proactive service identification and coexistence,'' in
	\emph{2020 54th Asilomar Conference on Signals, Systems, and Computers},
	2020, pp. 174--178.
	
	\bibitem{ViWi-BT}
	M.~Alrabeiah, J.~Booth, A.~Hredzak, and A.~Alkhateeb, ``{ViWi} vision-aided
	mmwave beam tracking: Dataset, task, and baseline solutions,'' \emph{arXiv
		preprint arXiv:2002.02445}, 2020.
	
	\bibitem{tx_id}
	G.~Charan and A.~Alkhateeb, ``User identification: The key enabler for
	multi-user vision-aided wireless communications,'' \emph{arXiv preprint
		arXiv:2210.15652}, 2022.
	
	\bibitem{charan2021c}
	G.~Charan, T.~Osman, A.~Hredzak, N.~Thawdar, and A.~Alkhateeb,
	``Vision-position multi-modal beam prediction using real millimeter wave
	datasets,'' in \emph{2022 IEEE Wireless Communications and Networking
		Conference (WCNC)}, 2022, pp. 2727--2731.
	
	\bibitem{imran2023environment}
	S.~Imran, G.~Charan, and A.~Alkhateeb, ``Environment semantic aided
	communication: A real world demonstration for beam prediction,'' in
	\emph{IEEE ICC Workshops}, 2023.
	
	\bibitem{charan2022drone}
	G.~Charan, A.~Hredzak, C.~Stoddard, B.~Berrey, M.~Seth, H.~Nunez, and
	A.~Alkhateeb, ``Towards real-world {6G} drone communication: Position and
	camera aided beam prediction,'' in \emph{GLOBECOM 2022-2022 IEEE Global
		Communications Conference}.\hskip 1em plus 0.5em minus 0.4em\relax IEEE,
	2022, pp. 2951--2956.
	
	\bibitem{demirhan2022radar}
	U.~Demirhan and A.~Alkhateeb, ``Radar aided {6G} beam prediction: Deep learning
	algorithms and real-world demonstration,'' in \emph{2022 IEEE Wireless
		Communications and Networking Conference (WCNC)}.\hskip 1em plus 0.5em minus
	0.4em\relax IEEE, 2022, pp. 2655--2660.
	
	\bibitem{10049816}
	------, ``Integrated sensing and communication for 6g: Ten key machine learning
	roles,'' \emph{IEEE Communications Magazine}, vol.~61, no.~5, pp. 113--119,
	2023.
	
	\bibitem{jiang2022lidar}
	S.~Jiang, G.~Charan, and A.~Alkhateeb, ``Lidar aided future beam prediction in
	real-world millimeter wave {V2I} communications,'' \emph{IEEE Wireless
		Communications Letters}, vol.~12, no.~2, pp. 212--216, 2022.
	
	\bibitem{Ying2020}
	Z.~Ying, H.~Yang, J.~Gao, and K.~Zheng, ``A new vision-aided beam prediction
	scheme for mmwave wireless communications,'' in \emph{2020 IEEE 6th
		International Conference on Computer and Communications (ICCC)}.\hskip 1em
	plus 0.5em minus 0.4em\relax IEEE, 2020, pp. 232--237.
	
	\bibitem{DetChPred}
	M.~Alrabeiah and A.~Alkhateeb, ``Deep learning for {TDD} and {FDD} massive
	mimo: Mapping channels in space and frequency,'' in \emph{2019 53rd Asilomar
		Conference on Signals, Systems, and Computers}, 2019, pp. 1465--1470.
	
	\bibitem{DeepSense}
	A.~Alkhateeb, G.~Charan, T.~Osman, A.~Hredzak, J.~Morais, U.~Demirhan, and
	N.~Srinivas, ``Deepsense {6G}: A large-scale real-world multi-modal sensing
	and communication dataset,'' \emph{IEEE Communications Magazine}, 2023.
	
	\bibitem{Wang2009}
	{Junyi Wang}, {Zhou Lan}, {Chang-woo Pyo}, T.~{Baykas}, {Chin-sean Sum}, M.~A.
	{Rahman}, {Jing Gao}, R.~{Funada}, F.~{Kojima}, H.~{Harada}, and S.~{Kato},
	``Beam codebook based beamforming protocol for multi-{G}bps millimeter-wave
	{WPAN} systems,'' \emph{IEEE Journal on Selected Areas in Communications},
	vol.~27, no.~8, pp. 1390--1399, 2009.
	
	\bibitem{Hur2013}
	S.~{Hur}, T.~{Kim}, D.~J. {Love}, J.~V. {Krogmeier}, T.~A. {Thomas}, and
	A.~{Ghosh}, ``Millimeter wave beamforming for wireless backhaul and access in
	small cell networks,'' \emph{IEEE Transactions on Communications}, vol.~61,
	no.~10, pp. 4391--4403, 2013.
	
	\bibitem{Sub6PredMmWave}
	M.~Alrabeiah and A.~Alkhateeb, ``Deep learning for mmwave beam and blockage
	prediction using sub-6 {GH}z channels,'' \emph{IEEE Transactions on
		Communications}, vol.~68, no.~9, pp. 5504--5518, 2020.
	
	\bibitem{Wang2019}
	Y.~{Wang}, A.~{Klautau}, M.~{Ribero}, A.~C.~K. {Soong}, and R.~W. {Heath},
	``Mmwave vehicular beam selection with situational awareness using machine
	learning,'' \emph{IEEE Access}, vol.~7, pp. 87\,479--87\,493, 2019.
	
	\bibitem{alkhateeb2023real}
	A.~Alkhateeb, S.~Jiang, and G.~Charan, ``Real-time digital twins: Vision and
	research directions for {6G} and beyond,'' \emph{IEEE Communications
		Magazine}, pp. 1--7, 2023.
	
	\bibitem{jiang2023digital}
	S.~Jiang and A.~Alkhateeb, ``Digital twin based beam prediction: Can we train
	in the digital world and deploy in reality?'' in \emph{IEEE ICC Workshops},
	2023.
	
	\bibitem{VAR}
	V.~M. {De Pinho}, M.~L.~R. {De Campos}, L.~U. {Garcia}, and D.~{Popescu},
	``Vision-aided radio: User identity match in radio and video domains using
	machine learning,'' \emph{IEEE Access}, vol.~8, pp. 209\,619--209\,629, 2020.
	
	\bibitem{Andrews2016}
	J.~G. Andrews, T.~Bai, M.~N. Kulkarni, A.~Alkhateeb, A.~K. Gupta, and R.~W.
	Heath, ``Modeling and analyzing millimeter wave cellular systems,''
	\emph{IEEE Transactions on Communications}, vol.~65, no.~1, pp. 403--430,
	2017.
	
	\bibitem{2021GC_BlkPred}
	G.~Charan, M.~Alrabeiah, and A.~Alkhateeb, ``Vision-aided dynamic blockage
	prediction for {6G} wireless communication networks,'' in \emph{2021 IEEE
		International Conference on Communications Workshops (ICC Workshops)}, 2021,
	pp. 1--6.
	
	\bibitem{CBLNNets}
	M.~Alrabeiah, Y.~Zhang, and A.~Alkhateeb, ``Neural networks based beam
	codebooks: Learning mmwave massive {MIMO} beams that adapt to deployment and
	hardware,'' \emph{arXiv preprint arXiv:2006.14501}, 2020.
	
	\bibitem{CBRL}
	Y.~Zhang, M.~Alrabeiah, and A.~Alkhateeb, ``Reinforcement learning of beam
	codebooks in millimeter wave and terahertz mimo systems,'' \emph{IEEE
		Transactions on Communications}, vol.~70, no.~2, pp. 904--919, 2022.
	
	\bibitem{DLBook}
	\BIBentryALTinterwordspacing
	I.~Goodfellow, Y.~Bengio, and A.~Courville, ``Deep learning,'' 2016, book in
	preparation for MIT Press. [Online]. Available:
	\url{http://www.deeplearningbook.org}
	\BIBentrySTDinterwordspacing
	
	\bibitem{Alkhateeb2015}
	A.~{Alkhateeb}, G.~{Leus}, and R.~W. {Heath}, ``Limited feedback hybrid
	precoding for multi-user millimeter wave systems,'' \emph{IEEE Transactions
		on Wireless Communications}, vol.~14, no.~11, pp. 6481--6494, 2015.
	
	\bibitem{PatternRecog}
	C.~M. Bishop, \emph{Pattern recognition and machine learning}.\hskip 1em plus
	0.5em minus 0.4em\relax springer, 2006.
	
	\bibitem{resnet}
	K.~He, X.~Zhang, S.~Ren, and J.~Sun, ``Deep residual learning for image
	recognition,'' in \emph{Proceedings of the IEEE conference on computer vision
		and pattern recognition}, 2016, pp. 770--778.
	
	\bibitem{He2017}
	K.~He, G.~Gkioxari, P.~Dollar, and R.~Girshick, ``Mask {R-CNN},'' in
	\emph{Proceedings of the IEEE International Conference on Computer Vision
		(ICCV)}, Oct 2017.
	
	\bibitem{Dai2016}
	J.~Dai, K.~He, and J.~Sun, ``Instance-aware semantic segmentation via
	multi-task network cascades,'' in \emph{Proceedings of the IEEE Conference on
		Computer Vision and Pattern Recognition (CVPR)}, June 2016.
	
	\bibitem{multi_var_poly_reg}
	\BIBentryALTinterwordspacing
	T.~K. Keyes, ``Applied regression analysis and multivariable methods,''
	\emph{Technometrics}, vol.~43, no.~1, pp. 101--101, 2001. [Online].
	Available: \url{https://doi.org/10.1198/tech.2001.s552}
	\BIBentrySTDinterwordspacing
	
\end{thebibliography}
\end{document}